\documentclass[a4paper,fleqn,usenatbib]{mnras}

\usepackage{newtxtext,newtxmath}
\usepackage[T1]{fontenc}
\usepackage{ae,aecompl}

\usepackage{graphicx}	% Including figure files
\usepackage{amsmath}	% Advanced maths commands
\usepackage{amssymb}	% Extra maths symbols

\usepackage{mathrsfs}
\usepackage[flushleft]{threeparttable}

\title[The accuracy of asteroseismology]{Establishing the accuracy of asteroseismic mass and radius estimates of giant stars. I. Three eclipsing systems at [Fe/H]$\sim-0.3$ and the need for a large high-precision sample.}

\author[K. Brogaard et al.]{K. Brogaard,$^{1,2}$\thanks{E-mail: kfb@phys.au.dk}
C. J. Hansen,$^{3}$
A. Miglio,$^{1,2}$
D. Slumstrup,$^{2}$
S. Frandsen,$^{2}$
\newauthor 
J. Jessen-Hansen,$^{2}$
M. N. Lund,$^{2,1}$
D. Bossini,$^{1}$
A. Thygesen,$^{4}$
G. R. Davies,$^{1,2}$
\newauthor 
W. J. Chaplin,$^{1,2}$
T. Arentoft,$^{2}$
H. Bruntt,$^{2}$
F. Grundahl,$^{2}$
and R. Handberg$^{2}$
\\
$^{1}$School of Physics and Astronomy, University of Birmingham, Edgbaston, Birmingham B15 2TT, UK\\
$^{2}$Stellar Astrophysics Centre, Department of Physics and Astronomy, Aarhus University, Ny Munkegade 120, 8000 Aarhus C, Denmark\\
$^{3}$Dark Cosmology Centre, The Niels Bohr Institute, Juliane Maries Vej 30, 2100, Copenhagen, Denmark\\
$^{4}$California Institute of Technology, 1200 E. California Blvd., MC 249-17, Pasadena, CA, 91125
}

\date{Accepted XXX. Received YYY; in original form ZZZ}

\pubyear{2017}

\begin{document}
\label{firstpage}
\pagerange{\pageref{firstpage}--\pageref{lastpage}}
\maketitle

\begin{abstract}
%This is a simple template for authors to write new MNRAS papers.
%The abstract should briefly describe the aims, methods, and main results of the paper.
%It should be a single paragraph not more than 250 words (200 words for Letters).
%No references should appear in the abstract.

We aim to establish and improve the accuracy level of asteroseismic estimates of mass, radius, and age of giant stars.
This can be achieved by measuring independent, accurate, and precise masses, radii, effective temperatures and metallicities of long period eclipsing binary stars with a red giant component that displays solar-like oscillations. 
%We measured precise properties of the three eclipsing binary systems KIC\,7037405, KIC\,9540226, and KIC\,9970396, finding for the giant components their masses to a precision of $1.7\%, 2.8\%$, and $1.3\%$, and their radii to a precision of $0.7\%, 1.2\%$, and $0.9\%$. Then, using log$g$ with the disentangled spectra we also determined $T_{\rm eff}$ and [Fe/H] and estimated the ages of the systems to be $5.3\pm0.5$, $3.1\pm0.6$, and $4.8\pm0.5$ Gyr for the adopted stellar model physics.
We measured precise properties of the three eclipsing binary systems KIC\,7037405, KIC\,9540226, and KIC\,9970396 and estimated their ages be $5.3\pm0.5$, $3.1\pm0.6$, and $4.8\pm0.5$ Gyr. The measurements of the giant stars were compared to corresponding measurements of mass, radius, and age using asteroseismic scaling relations and grid modeling. 
We found that asteroseismic scaling relations without corrections to $\Delta \nu$ systematically overestimate the masses of the three red giants by 11.7\%, 13.7\%, and 18.9\%, respectively. However, by applying theoretical correction factors $f_{\Delta \nu}$ according to \citet{Rodrigues2017}, we reached general agreement between dynamical and asteroseismic mass estimates, and no indications of systematic differences at the precision level of the asteroseismic measurements.
The larger sample investigated by \citet{Gaulme2016} showed a much more complicated situation, where some stars show agreement between the dynamical and corrected asteroseismic measures while others suggest significant overestimates of the asteroseismic measures. We found no simple explanation for this, but indications of several potential problems, some theoretical, others observational. Therefore, an extension of the present precision study to a larger sample of eclipsing systems is crucial for establishing and improving the accuracy of asteroseismology of giant stars.

\end{abstract}


%%%%%%%%%%%%%%%%%%%%%%%%%%%%%%%%%%%%%%%%%%%%%%%%%%

%%%%%%%%%%%%%%%%% BODY OF PAPER %%%%%%%%%%%%%%%%%%

\section{Introduction}

%This is a simple template for authors to write new MNRAS papers.
%See \texttt{mnras\_sample.tex} for a more complex example, and \texttt{mnras\_guide.tex}
%for a full user guide.

%All papers should start with an Introduction section, which sets the work
%in context, cites relevant earlier studies in the field by \citet{Others2013},
%and describes the problem the authors aim to solve \citep[e.g.][]{Author2012}.

Asteroseismology offers great prospects for new insights into stars, planets and our Galaxy through the exploitation of high-precision photometric time series from current and upcoming space missions. However, in order to ensure correct interpretation of the rapidly increasing amounts of observational data it is crucial that we establish the accuracy level of the asteroseismic methods.

The most easily extracted asteroseismic parameters are the frequency of maximum power, $\nu_{\mathrm{max}}$ and the large frequency spacing between modes of the same degree, $\Delta \nu$. $\Delta \nu$ has been shown to scale approximately with the mean density of a star \citep{Ulrich1986} while $\nu_{\mathrm{max}}$ scales approximately with the acoustic cut-off frequency of the atmosphere, which is related to surface gravity and effective temperature \citep{Brown1991,Kjeldsen1995,Belkacem2011}. In equation form these relations are:

\begin{eqnarray}\label{eq:01}
\frac{\Delta \nu}{\Delta \nu _{\odot}} & \simeq & f_{\Delta \nu}\left(\frac{\rho}{\rho_{\odot}}\right)^{1/2},\\
\label{eq:02}
\frac{\nu _{\mathrm{max}}}{\nu _{\mathrm{max,}\odot}} & \simeq & f_{\nu _{\mathrm{max}}} \frac{g}{g_{\odot}}\left(\frac{T_{\mathrm{eff}}}{T_{\mathrm{eff,}\odot}}\right)^{-1/2}.
\end{eqnarray}

Here, $\rho$, $g$, and $T_{\rm eff}$ are the mean density, surface gravity, and effective temperature, and we have adopted the notation of \citet{Sharma2016} that includes the correction functions $f_{\Delta \nu}$ and $f_{\nu _{\mathrm{max}}}$. By rearranging, expressions for the mass and radius can be obtained: 

\begin{eqnarray}\label{eq:03}
\frac{M}{\mathrm{M}_\odot} & \simeq & \left(\frac{\nu _{\mathrm{max}}}{f_{\nu _{\mathrm{max}}}\nu _{\mathrm{max,}\odot}}\right)^3 \left(\frac{\Delta \nu}{f_{\Delta \nu}\Delta \nu _{\odot}}\right)^{-4} \left(\frac{T_{\mathrm{eff}}}{T_{\mathrm{eff,}\odot}}\right)^{3/2},\\
\label{eq:04}
\frac{R}{\mathrm{R}_\odot} & \simeq & \left(\frac{\nu _{\mathrm{max}}}{f_{\nu _{\mathrm{max}}}\nu _{\mathrm{max,}\odot}}\right) \left(\frac{\Delta \nu}{f_{\Delta \nu}\Delta \nu _{\odot}}\right)^{-2} \left(\frac{T_{\mathrm{eff}}}{T_{\mathrm{eff,}\odot}}\right)^{1/2}. 
\end{eqnarray}

Although some empirical tests of these equations have been performed \citep{Brogaard2012, Miglio2012, Handberg2017, Huber2017}, a much larger effort is needed to establish the obtainable accuracy in general. Precise and accurate observations spanning a range in stellar parameters are needed because $f_{\Delta \nu}$, and potentially also $f_{\nu _{\mathrm{max}}}$, are non-linear functions of the stellar parameters. The solar reference values, which we adopt in this work to be $\Delta \nu _{\odot} = 134.9 \mu$Hz and $\nu _{\mathrm{max,}\odot} = 3090 \mu$Hz following \citet{Handberg2017} are also subject to uncertainties which further complicates tests of the correction factors.

The dependence of $f_{\Delta \nu}$ on stellar temperature, metallicity, and mass was first demonstrated by \citet{White2011} and later, in more detail and including core-helium-burning stars, by \citet{Miglio2013}, \citet{Sharma2016}, \citet{Rodrigues2017} and Serenelli et al. (in prep). \citet{Guggenberger2017} also published predictions for $f_{\Delta \nu}$, but not for the core-helium-burning phase. 
These articles provide figures, formulae, or codes that provide $f_{\Delta \nu}$ for a given combination of stellar parameters, which can be used with the scaling relations. 

The predicted $f_{\Delta \nu}$ are obtainable, because we understand $\Delta \nu$ well enough to derive the radial mode frequencies of a stellar model and compare the model $\Delta \nu$ to the mean density. However, when dealing with real stars, this is more complicated because in the general case we do not know the mass in advance and because errors in both the observed $T_{\rm eff}$ and the model temperature scale can introduce systematic errors in $f_{\Delta \nu}$. It is worth stressing that these complications are also present for asteroseismic grid modeling because they are not related to the scaling relations but rather the accuracy of the observables and stellar models. Additionally, due to the so-called surface effect, the measured $\Delta \nu$ of the Sun will be slightly different from that of a solar model ($\sim0.8$\%). While this was accounted for by \citet{White2011} and \citet{Rodrigues2017} by using the $\Delta \nu$ of the {\it model} Sun when calculating $f_{\Delta \nu}$, this was not done by \citet{Sharma2016} and therefore the corrections of the latter will not reproduce the parameters of the Sun. However, whether or not surface effects are in fact similar for the Sun and giant stars remains to be investigated.

We would like to point out that unlike the above mentioned $\Delta \nu$ corrections, the one proposed by \citet{Mosser2013} was not based on deviations from homology, but rather on errors caused by not being in the asymptotic frequency regime as assumed in the derivation of the $\Delta \nu$ scaling relation. The neglect of this should however not be of concern as long as the models are treated in exactly the same way as the observations when deriving $f_{\Delta \nu}$.

Unlike $\Delta \nu$, $\nu _{\mathrm{max}}$ is not yet understood to a level where it can be modeled directly although some efforts have been made to obtain a physical understanding \citep{Belkacem2011}. Very recently, \citet{Viani2017} have suggested that $f_{\nu _{\mathrm{max}}}$ might include the stellar mean molecular weight, $\mu$ and the adiabatic exponent, $\Gamma_1$ so that 
\begin{equation}
f_{\nu _{\mathrm{max}}}\simeq\left(\frac{\mu}{\mu_\odot}\right)^{1/2}\left(\frac{\Gamma_1}{\Gamma_{1,\odot}}\right)^{1/2},
\end{equation}
but no empirical tests have yet been made.

Eclipsing binaries are the only stars for which precise, accurate and model-independent radii \textit{and} masses can be measured. Aided by modern observational techniques and analysis methods these objects continue to allow stringent tests of stellar evolution theory and the asteroseismic methods. This is crucial for obtaining the precise and accurate age estimates of stars which are a requirement for the success of current and upcoming missions like {\it Kepler} \citep{Borucki2010}, K2 , TESS \citep{Ricker2014}, and PLATO \citep{Rauer2014}. 

\citet{Brogaard2016} described how eclipsing binary stars can be used for establishing the accuracy level of masses and radii of giant stars measured with asteroseismic methods. Studies of a few individual systems were carried out by \citet{Frandsen2013} and \citet{Rawls2016}, while \citet{Gaulme2016} published measurements of a larger sample of eclipsing systems.

In Sects.~\ref{sec:obs} -- ~\ref{sec:age} of this paper we present observations and precise measurements of masses, radii, effective temperatures and metallicities of stars in three specific eclipsing systems, all containing an oscillating red giant star and a main sequence (MS) or turn-off companion, and all with similar metallicity ([Fe/H]$\sim-0.3$). In Sect.~\ref{sec:compare} we compare these to asteroseismic predictions using scaling relations and grid modeling to establish the accuracy level at this metallicity. In Sect.~\ref{sec:gaulme} we include eclipsing systems from other studies in an attempt to generalize the results to other metallicities. Finally, in Sect.~\ref{sec:conclusion}, we summarise, conclude and outline future work crucial to establishing and improving the accuracy of asteroseismology of giant stars across all masses and metallicities.

\section{Observations and observables}
\label{sec:obs}

\subsection{Targets}

\begin{figure}
   \centering
   \includegraphics[width=9.0cm]{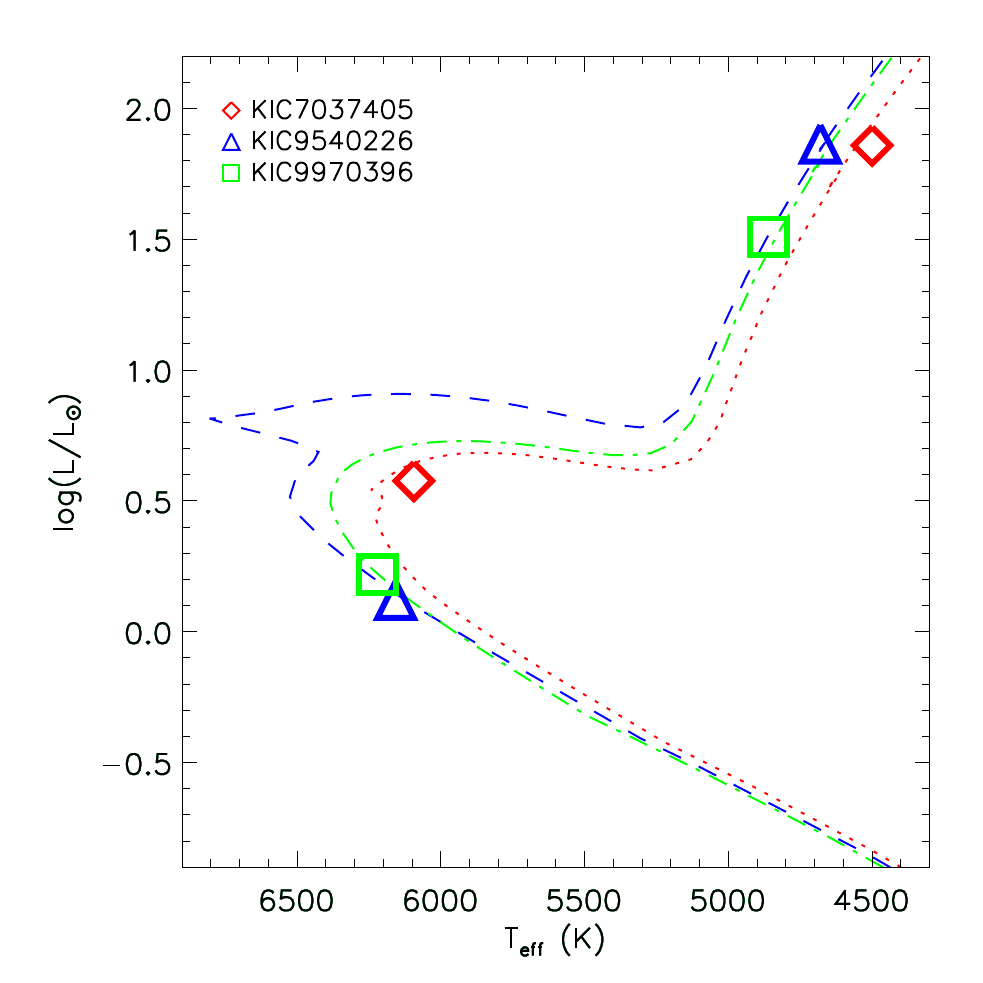}
   \caption{HR diagram showing the locations of the components of the eclipsing binaries KIC\,7037405, KIC\,9540226, and KIC\,9970396 and representative isochrones. }
             \label{fig:hrd}
    \end{figure}

Our targets are the eclipsing binaries KIC\,7037405, KIC\,9540226, and KIC\,9970396. They were selected from the {\it Kepler} eclipsing binary catalog \citep{Prsa2011} as targets showing solar-like oscillations and a total eclipse several days long. We aimed for those that were most likely to be SB2 binaries. 
During our spectroscopic follow-up observations these targets were also identified -- as part of a larger sample -- as eclipsing binaries with an oscillating red giant component by \citet{Gaulme2013} and measured by \citet{Gaulme2016}, although at lower spectral resolution and precision. Fig.~\ref{fig:hrd} shows the location of the components of each system in a Hertzsprung-Russell diagram along with representative isochrones based on the analysis presented in this paper (see Sect.~\ref{sec:age}).

%______________________________________________ 

\subsection{Photometry}

Light curves were constructed from \textit{Kepler} pixel data (Jenkins et al. 2010) downloaded from the KASOC database\footnote{\url{kasoc.phys.au.dk}} using the procedure
developed by S. Bloemen (private comm.) to automatically define pixel masks for aperture photometry. The extracted light curves were then corrected using the KASOC filter \citep{Handberg2014}. Briefly, the light curve is first corrected for jumps between observing quarters and is concatenated. It is then median filtered using two filters of different widths, to account for both spurious and secular variations, with the final filter being a weighted sum of the two filters based on the variability in the light curve. In addition to the median filters, the signal from the eclipses is iteratively estimated and included in the final filter from construction of the eclipse phase curve. This filtering allows one to isolate the different components of the light curve, and select which to be retained in the final light curve -- we refer to \citet{Handberg2014} for further details on the filtering methodology.
In the end, we only retained the long timescale and quarter-to-quarter adjustments. Finally, before the eclipse analysis we calculated the RMS of the light curve outside eclipses to be used as the photometric uncertainty and then removed most of the out-of-eclipse observations.

\subsection{Spectroscopy}

For spectroscopic follow-up observations we used the FIES spectrograph at the Nordic Optical Telescope and (for KIC\,9540226) also the HERMES spectrograph at the Mercator telescope, both located at the Observatorio del Roque de los Muchachos on La Palma. The FIES spectra were obtained using the HIGHRES setting which provides a resolution of $R\sim67000$ while the HERMES spectra have a resolution of $R\sim85000$. Table~\ref{tab:obs} gives the {\it Kepler} magnitude, the first and last observing dates, S/N, and number of observations for each target. Integration time with both instruments was 1800 seconds, with a few exceptions, e.g. interruption due to bad weather. Tables including the individual radial velocity measurements and the specific barycentric julian dates, calculated using the software by \citet{Eastman2010}, are given in the appendix.

\begin{table*}
\centering
\caption{Spectroscopic observation summary}
\label{tab:obs} 
\begin{tabular}{lcccclc}
\hline
Target & $K_p$ & First obs  & Last obs & Number of obs & Instrument & S/N@5606\AA (range,mean)\\
\hline
   KIC\,7037405 & 11.875 & $05-04-2014$ & $16-09-2016$ & 23 & FIES & 15-45, $29\pm6$\\
\hline
   KIC\,9540226 & 11.672 & $02-06-2011$ & $09-10-2011$ & 10 & FIES & 17-43, $22\pm7$ \\
   KIC\,9540226 &        & $22-07-2011$ & $14-08-2013$ & 32 & HERMES & 11-46, $21\pm7$ \\
\hline
   KIC\,9970396 & 11.447 & $29-04-2013$ & $05-08-2016$ & 22 & FIES & 29-56, $38\pm6$\\
\hline
\end{tabular}
\end{table*}

\subsubsection{Radial velocity measurements}

For measuring the radial velocities (RVs) of the binary components at each epoch, and to separate their spectra, we used a spectral separation code following closely the description of \citet{Gonzalez2006}. This is an iterative procedure where all the spectra are co-added at the radial velocity of one component and subtracted from each spectrum of the other, before measuring new best estimates of the radial velocities and iterating. For the radial velocity measurements we used the broadening function formalism by \citet{Rucinski1999,Rucinski2002} with synthetic spectra from the grid of \citet{Coelho2005}. 

This method works best when the observations sample a range in radial velocity as smoothly a possible. Four wavelength ranges were treated separately ($\lambda =$ 4500--5000, 5000--5500, 5500--5880, and 6000--6500 $\AA$). The gap between the two last wavelength ranges was introduced to avoid the region of the interstellar Na lines which causes problems for the spectral separation. For each epoch the final radial velocity was taken as the mean of the results from each wavelength range and the RMS scatter across wavelength was taken as a first estimate of the uncertainty. Later, when we fitted the binary solutions, we found that the first estimate RV uncertainties of the primary components were smaller than the RMS of the O-C of the best fit. This is not unexpected since our uncertainty estimate does not take into account uncertainty due to wavelength calibration errors and systematic uncertainty components such as scattered sunlight, artefacts in the spectra (e.g. due to cosmic rays and instrument imperfections) and imperfect subtraction of the secondary component spectral features. Therefore, a RV zero-point uncertainty of 140 $\rm m\,s^{-1}$ for KIC\,7037405 and KIC\,9970396, and 80 $\rm m\,s^{-1}$ for KIC\,9540226, was added in quadrature to the RV uncertainties in order for the analysis to yield a reduced $\chi^2$ close to 1 for the RV of both components of each system.

\subsubsection{Spectral analysis}
\label{sec:specanal}

The separated spectra of the giant components resulting from the above procedure were adjusted according to the light ratio of the eclipsing binary analysis (cf. Sect.~\ref{sec:binary}) to recover the true depths of the spectral lines. This was done to remove the continuum contribution from the secondary component. Specifically, this was achieved by multiplying the separated giant spectra by a factor of $\frac{L_{\rm G} + L_{\rm MS}}{L_{\rm G}}$ followed by a subtraction of $\frac{L_{\rm MS}}{L_{\rm G}}$. Here, $L_{\rm G}$ and $L_{\rm MS}$ refer to the luminosities of the giant and main-sequence star, respectively. We used the very precise light ratios from the $K_p$-band binary solution (cf. Sect.~\ref{sec:binary}) scaled to the $V$-band to be close to the wavelength range of the spectra, but as seen in Table~\ref{table:EBdata} these are very similar.

The stellar parameters were first derived using only the stellar spectra employing IRAF and MOOG spectrum analysis code \citep[][version 2014]{Sneden1973} together with Kurucz-type Atlas 9 models with solar-scaled opacity distribution functions \citep[][without convective overshooting]{Castelli2003}.

For this type of analysis, the equivalent widths of Fe lines are used by enforcing balances to determine the stellar parameters. Neutral Fe lines are temperature sensitive, and requiring that all Fe lines regardless of excitation potential yield the same abundance is one way of deriving temperatures (excitation balance). Similarly, ionised Fe lines are pressure and therefore gravity sensitive, and hence forcing ionisation balance $\rm{A(FeI) = A(FeII)}$ sets log$g$ of the model atmosphere. The microturbulence can be derived by requiring that all Fe lines (regardless of their strength) yield the same abundance thereby linking the microturbulence parameter and the metallicity.

By using equivalent widths and balances to derive stellar parameters these become interdependent. Furthermore, if the Fe lines are so strong that they saturate, they fall on the saturated or even damped part of the curve of growth, and the linear relation between equivalent width and abundance breaks down. If the lines are only mildly satured, a higher microturbulence can help delay saturation as the small scale velocities broaden the lines and affect the opacities, source function, and in turn the abundances (see fig. 16.5 in \citealt{Gray2005}). 

To avoid such potential bias we adopt the precise surface gravities derived from the dynamical solution of our radial velocity and eclipse modeling. The microturbulence was calculated using the Gaia-ESO empirical formula (Kovalev et al. 2018, in prep.), which relates the microturbulence to linear and quadratic terms in temperature, gravity, and metallicity, and it yields realistic velocity values (see Table \ref{table:stelpar}). This leaves us with optimising and deriving metallicity and temperature using equivalent widths.

%The temperature and metallicity were iterated and adjusted until equilibrium occurred.

\begin{table*}
\centering
\caption{\label{table:stelpar}Atmospheric parameters for the programme stars}
\begin{tabular}{lcclccc}
\hline
Star & $T_{\rm eff}$  & log$g$ & log$\epsilon(\rm{Fe_I} / \rm{Fe_{II}})$ & (No. of $\rm{Fe_I}$ / $\rm{Fe_{II}}$ lines) & $\rm{[Fe/H]}$ & $\xi$ \\
  & [K] & [cgs]  & [dex] &  & [dex] & [km$\times$s$^{-1}$]\\
\hline
   7037405A &  $4500\pm50$    &   $2.22$  & $7.30\pm0.02/7.23\pm0.04$ & (37,16) &  $-0.27$   & $1.4\pm0.1$ \\
   9540226A &  $4680\pm50$    &   $2.35$  & $7.25\pm0.02/7.27\pm0.03$ & (67,22) &  $-0.23$   & $1.3\pm0.1$ \\
   9970396A &  $4860\pm30$    &   $2.70$  & $7.30\pm0.02/7.15\pm0.03$ & (79,24) &  $-0.35$   & $1.2\pm0.1$ \\
\hline
\end{tabular}
\end{table*}

In the spectral regions 4500 -- 5880\,$\AA$ and 6000 -- 6500\,$\AA$ we considered 106 FeI and 30 FeII lines using the line list employed in \citet{Hansen2012}. However, many of the bluest lines are heavily blended reducing the number of lines useful for abundance analysis to around 80 Fe I lines and 20 Fe II lines. Below 4500\,$\AA$ the line density is so large that lines from various species all blend into one line profile, e.g., that of the Fe lines we wish to measure. Luckily, this is typically expressed in a larger profile width. Instrumental broadening is folded into the overall Gaussian line width as the instrumental resolution enters the Gaussian through the full width half maximum (FWHM). Lines with too large FWHM compared to the instrumental profile or equivalent widths larger than 250m$\AA$ were also rejected from the analysis as these were either blended or saturated, this reduced the number of lines further. The final number of lines used to derive temperature and metallicity are listed in Table \ref{table:stelpar} along with the effective temperatures and their uncertainties estimated from the fitted slopes folded with the number of Fe lines. For the metallicities the standard deviation on the mean is given. From Table \ref{table:stelpar} $\rm{FeI}$ and $\rm{FeII}$ are seen to provide ionisation equilibrium within $1\sigma$ only for KIC\,9540226A, while being at the limit of the $1\sigma$ level for KIC\,7037405A, and close to $3\sigma$ for KIC\,9970396A. An independent analysis using another line list, MARCS atmosphere models and astrophysical oscillator strengths calibrated on the Sun reproduced our results well within errors, including the ionisation non-equilibrium. Similar issues have been reported by e.g. \citet{Morel2014} and \citet{Rawls2016} for optical spectra and by \citet{Pinsonneault2014} and \citet{Holtzman2015} for APOGEE H-band spectra. 
Our analysis assumes local thermodynamic equilibrium (LTE), where the energy transport is collision-dominated. This is close to reality in the innermost parts of the atmosphere where FeII lines form. This is however, a poorer assumption for the outer layers where FeI lines typically form and where radiation plays a larger role. We therefore adopt values for [Fe/H] based on $\rm{FeII}$ to reduce potential effects caused by the departure from LTE. In Sect.~\ref{sec:age} we find some support for this choice.

As a final test of the quality of the derived stellar parameters, lines in the regions 5700-5800\,$\AA$\, and 6100-6200\,$\AA$\, were fitted with synthetic spectra adopting a model with the best set of parameters. The synthetic spectra fit all Fe lines in these regions to within $\pm 0.1$\,dex indicating that the adopted gravity, and derived metallicity and temperature are reasonable. Our derived effective temperatures agree within the errors with the independent analysis by \citet{Gaulme2016}.
The uncertainties are only reflecting internal errors of the procedures and do not account for correlations between parameters. Moreover, they reflect precision rather than accuracy. We therefore rely on the results of a detailed investigation by \citet{Bruntt2010} for adopting total uncertainties of 80 K for $T_{\rm eff}$ and 0.1 dex for [Fe/H] to allow for systematics when comparing to models later, and experiment also with using either FeI or FeII for our [Fe/H] estimate. 

\section{Eclipsing binary analysis}
\label{sec:binary}
To determine model independent stellar parameters we used the JKTEBOP eclipsing binary code \citep{Southworth2004} which is based on the EBOP program developed by P. Etzel \citep{Etzel1981,Popper1981}. We made use of several features of the program that have been developed later. The most important are non-linear limb darkening \citep{Southworth2007}, simultaneous fitting of the light curve and the measured radial velocities \citep{Southworth2013}, and numerical integration \citep{Southworth2011}. The latter is needed due to the long integration time of {\it Kepler} long cadence photometry (24.9 minutes).

We fitted for the following parameters : Orbital period $P$, first eclipse of the giant component $T_G$, surface brightness ratio $J$, sum of the relative radii $r_{\rm MS}+r_{\rm G}$, ratio of the radii $k=\frac{r_{\rm MS}}{r_{\rm G}}$, orbit inclination $i$, $e$cos$\omega$, $e$sin$\omega$, semi-amplitudes of the components $K_{\rm G}$ and $K_{\rm MS}$ and system velocity of the components $\gamma_{\rm G}$ and $\gamma_{\rm MS}$. We allow for two system velocities because the components and their analysis are affected differently by gravitational redshift \citep{Einstein1952} and convective blueshift \citep{Gray2009} effects.

We used a quadratic limb darkening law with coefficients calculated using JKTLD \citep{Southworth2015} with tabulations for the $K_p$ bandpass by \citet{Sing2010}. We ran JKTEBOP iteratively, starting with limb darkening coefficients from first guesses and then using $T_{\rm eff}$ for the red giant from the spectral analysis with log$g$ fixed from the binary solution. A $T_{\rm eff}$ estimate for the main sequence component was obtained by reproducing the light ratio for the $Kp$ passband from JKTEBOP using Planck functions modified according to the ratio of the radii $k$. New limb darkening coefficients were then calculated with JKTLD using these $T_{\rm eff}$ and log$g$ values to be used in the next JKTEBOP solution.

Gravity darkening coefficients were taken from \citet{Claret2011} though large changes to these numbers had negligible effects as expected for nearly spherical stars. The same is true for reflection effects, which were calculated from system geometry. Light contamination from other stars was treated as third light and was fixed to the mean contamination of quarters 1-17 given on the \textit{Kepler} MAST web-page\footnote{https://archive.stsci.edu/kepler/data\_search/search.php}, which is 0.020, 0.000, and 0.003 for KIC\,7037405, KIC\,9540226, and KIC\,9970396, respectively.  

\begin{figure}
   \centering
   \includegraphics[width=8.6cm]{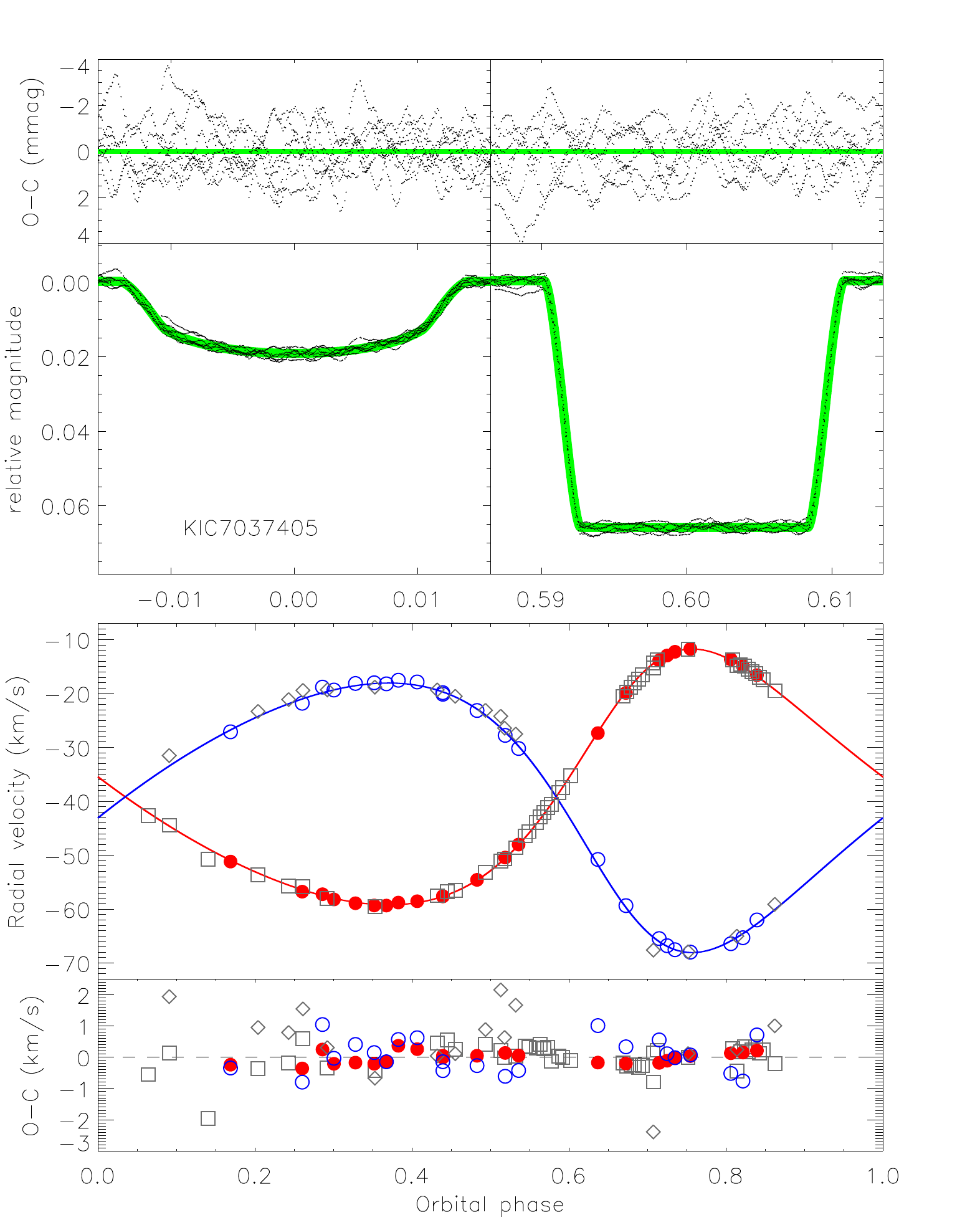}
   \caption{Binary Model fit to {\it Kepler} light curve (upper panels) and radial velocities (lower panels) for KIC\,7037405. Red indicates the giant component, blue the main sequence component. Filled and open circles represent our radial velocity measurements of the giant and main sequence star, respectively. Grey squares and diamonds represent the measurements of \citet{Gaulme2016} for the giant and main sequence star, respectively.}
             \label{fig:7037405}%
    \end{figure}

\begin{figure}
   \centering
   \includegraphics[width=8.6cm]{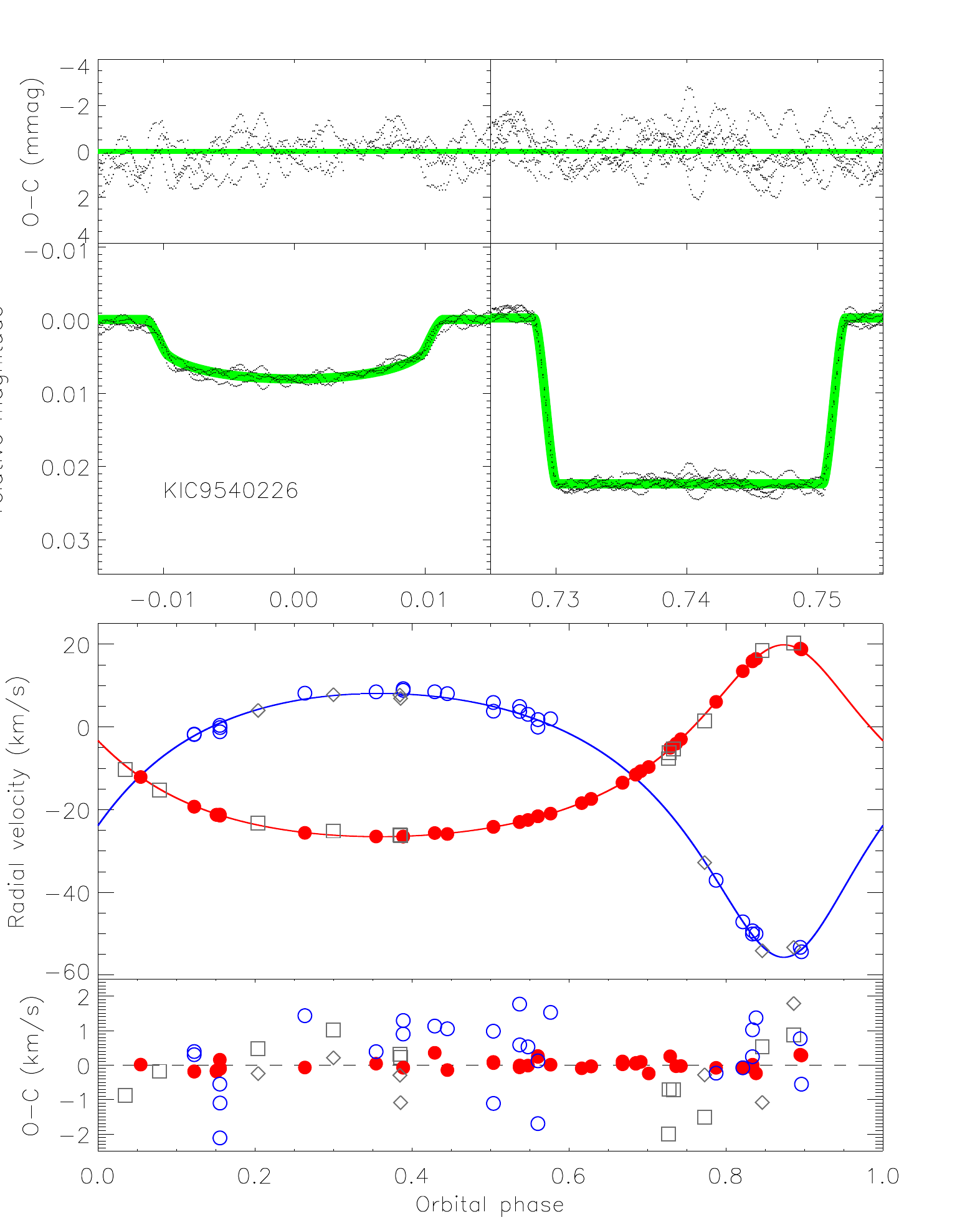}
   \caption{As Fig.~\ref{fig:7037405}, but for KIC\,9540226.}
             \label{fig:9540226}%
    \end{figure}

\begin{figure}
   \centering
   \includegraphics[width=8.6cm]{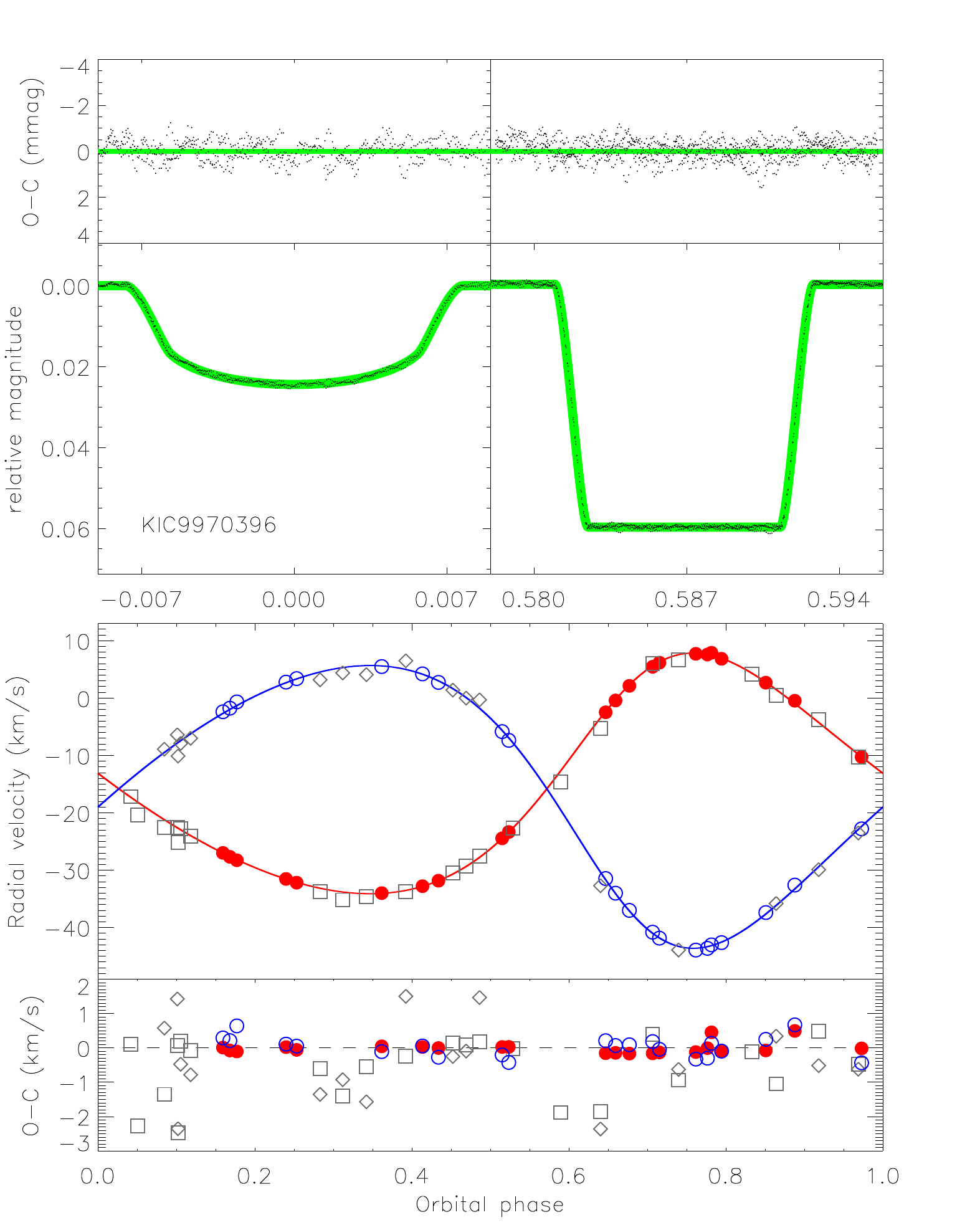}
   \caption{As Fig.~\ref{fig:7037405}, but for KIC\,9970396.}
             \label{fig:9970396}%
    \end{figure}

The optimal JKTEBOP solutions are compared to the observed light curves and measured radial velocities in Figs.~\ref{fig:7037405} --~\ref{fig:9970396}. It is clear from the light curve O-C diagram, that the residuals are dominated by the solar-like oscillations rather than random errors. We therefore used the residual-permutation uncertainty estimation method of JKTEBOP which accounts for correlated noise when estimating parameter uncertainties. The final parameters and their uncertainties are given in Table~\ref{table:EBdata}.

Our mass estimates are for all giants different by close to or more than the one-sigma limits of the corresponding measurements by \citet{Gaulme2016}. Specifically, our mass values are different from those of \citet{Gaulme2016} by $-2\sigma$, $1.25\sigma$, and $0.96\sigma$ (their uncertainties) for KIC\,7037405A, KIC\,9970396A, and KIC\,9540226A, respectively. 

The RV O-C diagrams show our residuals as well as those of \citet{Gaulme2016} when their RV measurements are phased to our solution. In general our results are more precise, which is not surprising given the higher resolution of our spectra. However, a comparison of only the RVs of the primary components seems to indicate that the study by \citet{Gaulme2016} suffers from epoch to epoch RV zero-point issues because the (O-C) values of their measurements are much larger than the claimed RV uncertainties. Only the RVs for the secondary component of KIC\,9540226 are the (O-C)s by \citet{Gaulme2016} comparable to ours. Indeed, this system has the smallest light ratio of the three. The main sequence star only contributes 2\% of the light, 3-4 times less than for the other two systems. This makes the RV measurements much more sensitive to potential systematic error sources such as scattered sunlight and incomplete subtraction of the giant component spectral lines in the spectral separation process. Therefore, we manually inspected the broadening functions for the main sequence star and disregarded the RV measurement in cases where the broadening function looked significantly asymmetric or could not be reliably identified. This is the main reason that our mass and radius estimates for KIC\,9540226 deviates from our preliminary result in \citet{Brogaard2016} where in hindsight the estimated 2\% uncertainty on mass and radius was too optimistic. The lack of manual inspection of the broadening functions in that preliminary analysis caused the inclusion of radial velocities for spectra that did not show a broadening function peak, likely due to too low S/N. Those radial velocitiy measurements were therefore not caused by signal, but by noise in the broadening function. We note that Gaulme et al. (2016) must have somehow made a similar kind of selection, since they do also not measure the radial velocity of the secondary star at all epochs for this star.

With more observations the mass and radius uncertainty for KIC\,9540226, as well as the other systems can be further reduced. Not just because more measurements reduces the random error of the binary solution, but also because each spectrum contributes to the combined component spectra that are subtracted from each of the individual spectra to calculate RVs in the spectral separation algorithm. Therefore, each observed spectrum increases the S/N of the RV calculation of all RVs of the given system. With enough observations the combined separated spectrum of the main sequence components could reach a high enough S/N that they could also be used for $T_{\rm eff}$ and metallicity measurements. This could potentially reveal atomic diffusion signals through a comparison of the element abundances between the main sequence and giant component; Atomic diffusion will cause heavy elements sink into a star during the main sequence evolution, while they return to the surface once the star becomes a giant because of the deep convection zone that develops. Thus, the main sequence star should have slightly different surface abundances than the giant star. 

\begin{table*}
\centering
\caption{\label{table:EBdata}Properties of the eclipsing binaries.}
\begin{threeparttable}
\begin{tabular}{lccc}
\hline
\hline
Quantity & KIC\,7037405 & KIC\,9540226 & KIC\,9970396 \\
\hline
RA  (J2000)\tnote{9}                                & $19:31:54.293$   & $19:48:08.158$   & $19:54:50.352$ \\
DEC (J2000)\tnote{9}                                & $+42:32:51.65$   & $+46:11:54.49$   & $+46:49:58.91$ \\
$K_p$		                                    &    11.875        &    11.672        & 11.447            \\ 
% $V$		                                    &                  &                  &              \\ 	
\hline
Orbital period (days)                               & $207.10849(95)$  & $175.44301(36)$  & $235.29861(24)$ \\
Reference time $\rm T_{\rm RG}$                     & $54988.3929(83)$ & $55161.096(20)$  & $55052.4477(35)$ \\
Inclination $i$ ($\circ$)                           & $88.469(62)$     &   $89.43(0.37)$  & $89.437(46)$     \\
Eccentricity $e$ ($\circ$)                          & $0.2364(26)$     & $0.38782(24)$    & $0.1942(53)$    \\
Periastron longitude $\omega$                       & $311.30(59)$     & $3.32(38)$       & $314.2(15)$    \\
Sum of the fractional radii $r_{\rm MS}+r_{\rm RG}$ & $0.08128(49)$    & $0.07988(77)$    & $0.04405(45)$\\
Ratio of the radii $k$                              & $0.12465(38)$    & $0.07763(71)$    & $0.13799(85)$\\
Surface brightness ratio $J$                        & $3.672(25)$      & $3.175(68)$      & $2.746(19)$\\
$\frac{L_{\rm MS}}{L_{\rm RG}} K_p$                 & $0.06219(13)$    & $0.020607(98)$   & $0.056344(53)$\\
$K_{\rm RG}$                                        & $23.728(49)$     & $23.191(49)$     & $20.971(58)$ \\
$K_{\rm MS}$                                        & $25.00(21)$      & $31.90(40)$      & $24.64(14)$  \\
semi-major axis $a (\rm R_{\odot})$                   & $193.80(89)$     & $176.0(13)$      & $207.92(73)$ \\
$\gamma_{\rm RG}$                                   & $-39.176(14)$    & $-12.323(11)$    & $-15.978(16)$ \\
$\gamma_{\rm MS}$                                   & $-39.10(13)$     & $-11.13(25)$     & $-15.592(55)$ \\
Mass$_{\rm RG}(M_{\odot})$                          & $1.170(20)$      & $1.378(38)$      & $1.178(15)$\\
Mass$_{\rm MS}(M_{\odot})$                          & $1.110(11)$      & $1.002(15)$      & $1.0030(85)$\\
Radius$_{\rm RG}(M_{\odot})$                        & $14.000(93)$     & $13.06(16)$      & $8.035(74)$\\
Radius$_{\rm MS}(M_{\odot})$                        & $1.746(14)$      & $1.014(14)$      & $1.1089(52)$\\
log$g_{\rm RG}$ (cgs)                               & $2.2131(67)$     & $2.345(10)$      & $2.699(11)$ \\
log$g_{\rm MS}$ (cgs)                               & $3.9990(71)$     & $4.427(10)$      & $4.3493(54)$ \\
\hline
$\frac{L_{\rm MS}}{L_{\rm RG}}$ bolometric          & $0.0557(8)$      & $0.0190(4)$      & 0.0534(7)             \\
$\frac{L_{\rm MS}}{L_{\rm RG}} V$                   & $0.0738(3)$      & $0.02387(3)$     & 0.0638(3)             \\
$T_{\rm eff,RG}$                                    & $4500\pm80$      & $4680\pm80$      & $4860\pm80$ \\
$T_{\rm eff,MS}$                                    & $6094\pm138$     & $6157\pm131$     & $6221\pm125$   \\
$\rm [FeI/H]$	                                    & $-0.20\pm0.02$   & $-0.25\pm0.02$	  & $-0.20\pm0.02$	\\
$\rm [FeII/H]$	                                    & $-0.27\pm0.05$   & $-0.23\pm0.03$	  & $-0.35\pm0.03$	\\
$\rm[Fe/H]$	                                    & $-0.27\pm0.10$   & $-0.23\pm0.10$	  & $-0.35\pm0.10$	\\
\hline
\end{tabular}

\begin{tablenotes}
		\scriptsize
\item[9] From the KIC.

\end{tablenotes}
\end{threeparttable}
\end{table*}

\begin{table*}
\centering
\caption{\label{table:RGdata}Measurements of the red giants.}
\begin{threeparttable}
\begin{tabular}{lccc}
\hline\hline
Quantity & KIC\,7037405A & KIC\,9970396A & KIC\,9540226A\\
\hline
$\nu_{\rm max}$\tnote{1} & $21.75\pm0.14$   & $63.70\pm0.16$  & $27.07\pm0.15$  \\
$\Delta\nu$\tnote{1} & $2.792\pm0.012$  & $6.320\pm0.010$ & $3.216\pm0.013$ \\

$f_{\Delta\nu}$ correction factor\tnote{2}  & 0.964       &   0.970        &   0.967     \\
$f_{\nu_{\rm max}}$ from mass               & 0.988       &   1.018        &   0.987     \\
$f_{\nu_{\rm max}}$ from radius             & 0.963       &   1.009        &   0.993     \\

%seismic masses and radii and correction factors are calculated using seis_binaries.pro
\hline
%Mass$_{\rm dyn}(\rm M_{\odot})$                  & $1.170(20)$               & $1.178(15)$               & $1.378(38)$\\
Mass$_{\rm dyn}(\rm M_{\odot})$                  & $1.170 \pm0.020$      & $1.178\pm0.015$               & $1.378\pm0.038$\\
Mass$_{\rm seis-raw}(\rm M_{\odot})$             & $1.307\pm0.049$           & $1.40\pm0.037$            &  $1.567\pm0.046$    \\
Mass$_{\rm seis-corr}(\rm M_{\odot})$\tnote{2}   & $1.128\pm0.042$           & $1.242\pm0.033$           &  $1.370\pm0.040$     \\
Mass$_{\rm PARAM,[FeII/H]}$                          & $1.156^{+0.040}_{-0.045}$ & $1.264^{+0.034}_{-0.017}$ & $1.309^{+0.048}_{-0.044}$ \\
Mass$_{\rm PARAM,[FeI/H]}$                           & $1.127^{+0.042}_{-0.036}$ & $1.204^{+0.020}_{-0.016}$ & $1.329^{+0.043}_{-0.048}$ \\
\hline

%Radius$_{\rm dyn}(\rm R_{\odot})$                & $14.000(93)$            &   $8.035(74)$          & $13.06(16)$  \\
Radius$_{\rm dyn}(\rm R_{\odot})$                & $14.000\pm0.093$            &   $8.035\pm0.074$          & $13.06\pm0.16$  \\
Radius$_{\rm seis-raw}(\rm R_{\odot})$           & $14.50\pm0.20$          &   $8.614\pm0.079$      &  $14.02\pm0.16$      \\
Radius$_{\rm seis-corr}(\rm R_{\odot})$\tnote{2} & $13.48\pm0.19$          &   $8.106\pm0.074$      &  $13.11\pm0.15$      \\
Radius$_{\rm PARAM,[FeII/H]}$              & $13.46^{+0.20}_{-0.23}$ & $8.10^{+0.09}_{-0.07}$ & $12.79^{+0.21}_{-0.20}$ \\
Radius$_{\rm PARAM,[FeI/H]}$               & $13.31^{+0.21}_{-0.18}$ & $7.94^{+0.07}_{-0.06}$ & $12.87^{+0.19}_{-0.20}$ \\
\hline

Age$_{\rm PARAM,[FeII/H]}$ (Gyr)    & $5.38^{+1.03}_{-0.70}$ & $3.64^{+0.16}_{-0.30}$ & $3.58^{+0.50}_{-0.48}$ \\
Age$_{\rm PARAM,[FeI/H]}$ (Gyr)     & $6.28^{+0.89}_{-0.91}$ & $4.86^{+0.30}_{-0.38}$ & $3.32^{+0.54}_{-0.38}$ \\
Age$_{\rm dyn,FeII}$ (Gyr)          & $5.4\pm0.5$            & $4.8\pm0.5$            & $3.1\pm0.6$ \\
Age$_{\rm{dyn},T_{\rm{eff}}}$ (Gyr) & $5.8\pm0.5$            & $4.6\pm0.5$            & $2.9\pm0.6$ \\
Age$_{\rm dyn,FeI}$ (Gyr)           & $5.7\pm0.5$            & $5.4\pm0.5$            & $3.0\pm0.6$ \\

%  7.03740e+06      5.37830      1.03270     0.700700      1.15640    0.0404000    0.0451000      13.4589     0.204700     0.227700
%  9.97040e+06      3.64200     0.159700     0.299800      1.26420    0.0337000    0.0166000      8.10300    0.0880003    0.0679998
%  9.54023e+06      3.58330     0.495400     0.481400      1.30930    0.0481000    0.0439999      12.7904     0.205200     0.195000
%  7.03740e+06      6.27730     0.888900     0.906100      1.12650    0.0418000    0.0366000      13.3106     0.211500     0.183101
% 9.97040e+06      4.85670     0.296300     0.377300      1.20360    0.0203999    0.0160000      7.93820    0.0656004    0.0612998
% 9.54023e+06      3.32770     0.535300     0.382800      1.32850    0.0439000    0.0477000      12.8704     0.194699     0.203300
\hline
\end{tabular}

\begin{tablenotes}
		\scriptsize
\item[1] Adopted from \citet{Gaulme2016}.
\item[2] Correction to $\Delta\nu$ according to \citet{Rodrigues2017} assuming RGB stars with [Fe/H]=-0.25 and the dynamical masses.

\end{tablenotes}
\end{threeparttable}
\end{table*}

%______________________________________________ 

%______________________________________________ 

%______________________________________________ 

\section{Dynamical age estimates}
\label{sec:age}

\begin{figure}
   \centering
   \includegraphics[width=8.6cm]{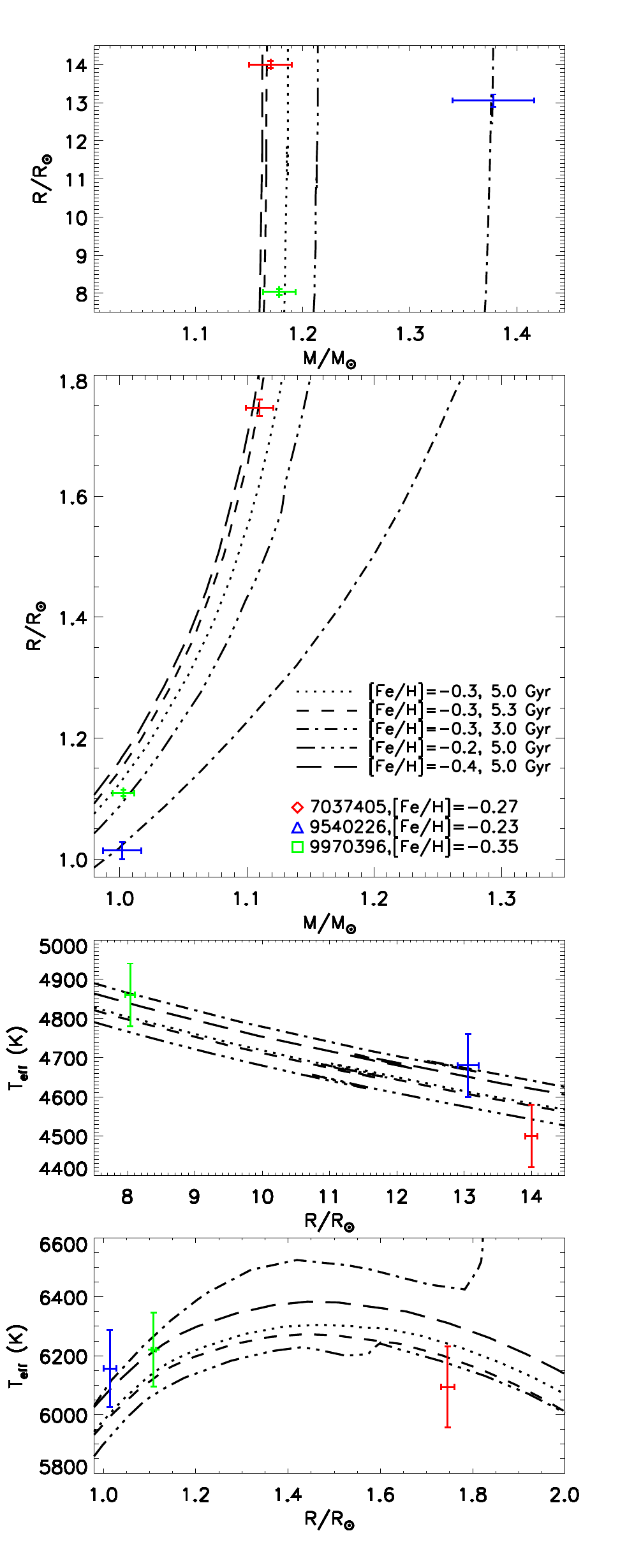}
   \caption{Mass--radius and radius--$T_{\rm eff}$ diagrams with measurements of the eclipsing binary components compared to isochrones of different ages and metallicities as indicated by the legend in the second panel.}
             \label{fig:mrt}
    \end{figure}

The ages of the stars can be estimated by comparing our measurements to isochrones. We adopted the isochrones used in the PARAM grid described by \citet{Rodrigues2017} in order to be able to do a direct comparison with ages derived from grid based asteroseismic modeling using PARAM with the observed $\Delta\nu$, $\nu_{\rm max}$, $T_{\rm eff}$, and [Fe/H]. Fig.~\ref{fig:mrt} shows mass--radius and radius--$T_{\rm eff}$ diagrams with our measurements compared to isochrones of different ages and metallicities. From such comparisons one can infer age estimates as well as corresponding uncertainties. The isochrones shown illustrate that the uncertainty in age due to uncertainty on radius estimates is negligible because the isochrones are essentially vertical at the location of the giant components in the mass--radius diagram, and thus the age uncertainty is almost entirely due to mass uncertainty for a fixed metallicity. The uncertainty in age due to uncertainty in mass is about $\pm0.3$ Gyr, which can be seen by comparing the dotted and short-dashed isochrones to the mass uncertainty of the giant components of KIC\,7037405 and KIC\,9970396 in the top panel of Fig.~\ref{fig:mrt}. Correspondingly, the uncertainty in age due to metallicity is $\pm0.4$ Gyr per $\pm0.1$ [Fe/H] as seen by including the long-dashed isochrone in the comparisons. If we conservatively adopt a $\pm0.1$ dex uncertainty on [Fe/H] then the total age uncertainty is $\pm0.5$ Gyr for this particular model grid when adding the contributions from mass and [Fe/H] in quadrature. We have not attempted to reduce the age uncertainty by including the secondary components in the analysis. This choice was made due to the correlation between the component masses which is present in eclipsing binary measurements.

Two of the systems appear to be close in age. If using only the mass-radius diagram and the specific [Fe/H] (based on FeII lines) measured for each system, KIC\,9970396 is $4.8\pm0.5$ Gyr and KIC\,7037405 is $5.4\pm0.5$ Gyr. For these systems, the radius--$T_{\rm eff}$ diagram supports that they actually have different [Fe/H], since that provides a better match between isochrones and observations. This lends some support to the choice of using the FeII lines for the [Fe/H] measurements in Sect.~\ref{sec:specanal}, given that the FeI lines suggested identical [Fe/H] for the two systems. In fact, if we chose to match the exact measurements of $T_{\rm eff}$ while adjusting instead [Fe/H], we would obtain a larger difference in [Fe/H] and ages of $4.6\pm0.5$ Gyr and $5.8\pm0.5$ Gyr for KIC\,9970396 and KIC\,7037405, respectively. However, the comparison of dynamical and asteroseismic masses below might indicate that the real $T_{\rm eff}$ difference between these systems is smaller than measured, in which case they might be very close to co-eval with nearly identical [Fe/H] as well, in agreement with what was found from the FeI lines. We give these three different age estimates in Table ~\ref{table:RGdata} along with the best age estimates of KIC\,9540226. This system is younger than the other two systems, $3.1\pm0.6$ Gyr at the measured metallicity where there is also close agreement with the measured $T_{\rm eff}$. 

%______________________________________________ 

\section{The accuracy of asteroseismic estimates of mass, radius, and age}
\label{sec:compare}
\begin{figure}
   \centering
   \includegraphics[width=8.0cm]{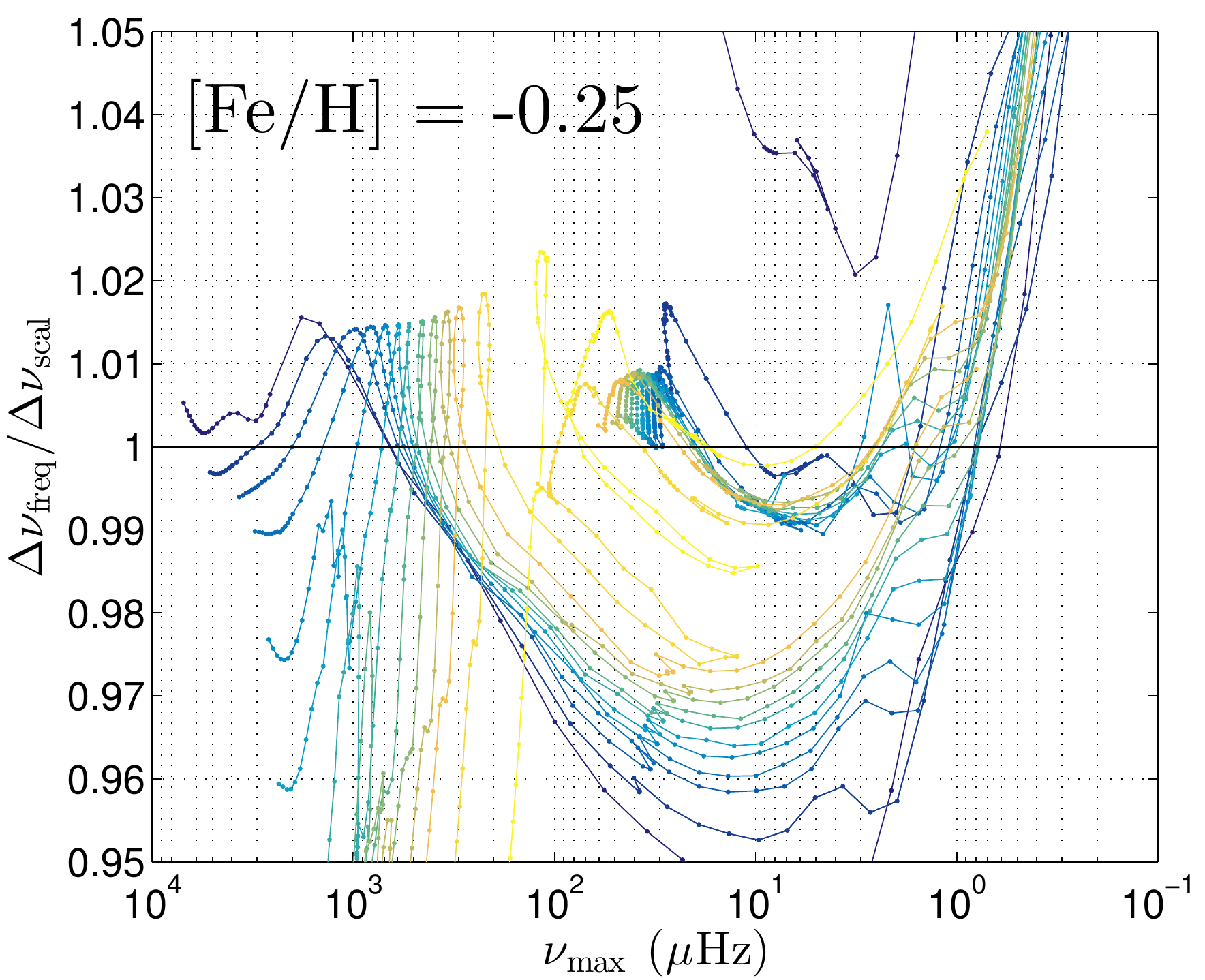}   
   \includegraphics[width=8.0cm]{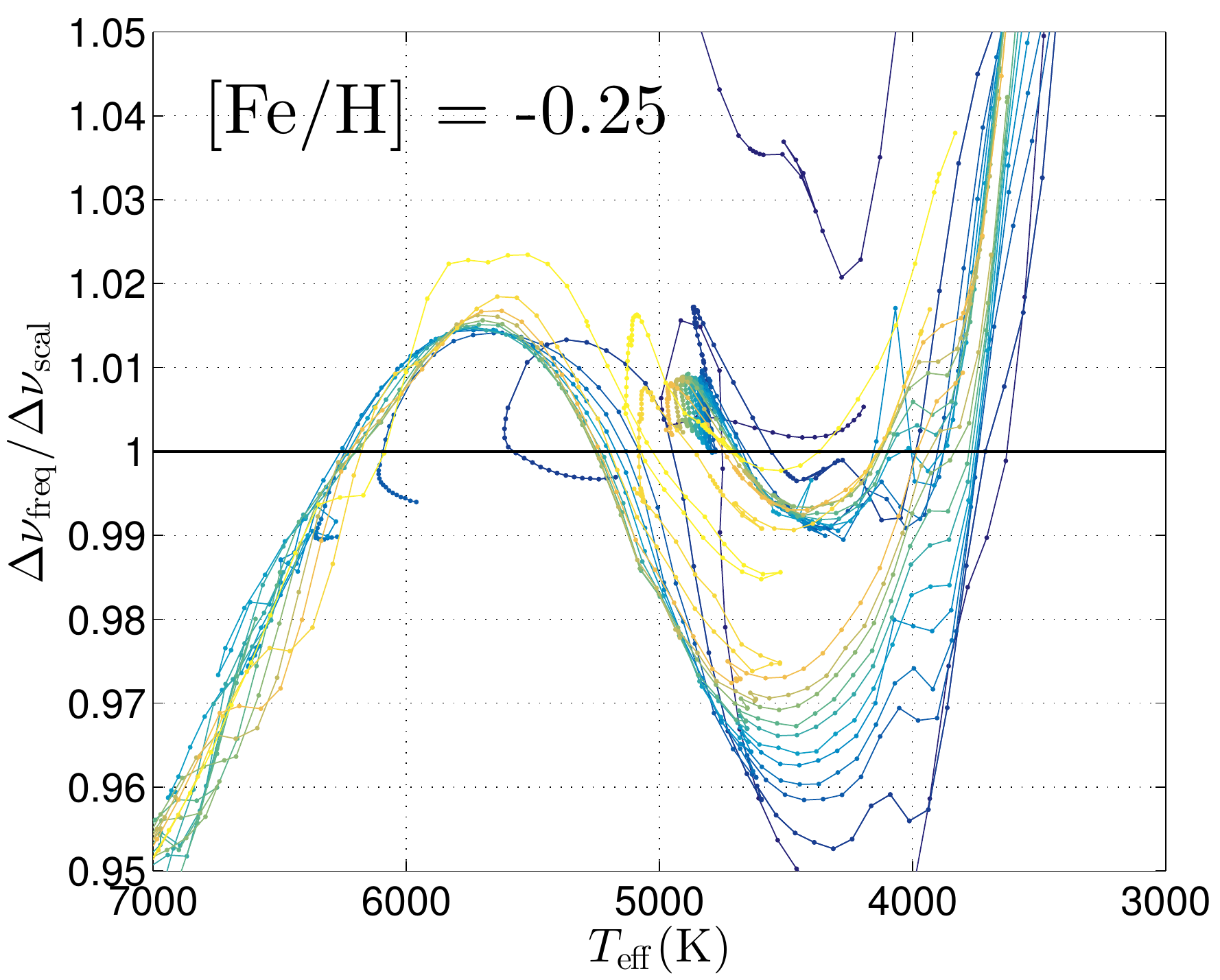}
   \includegraphics[width=4.0cm]{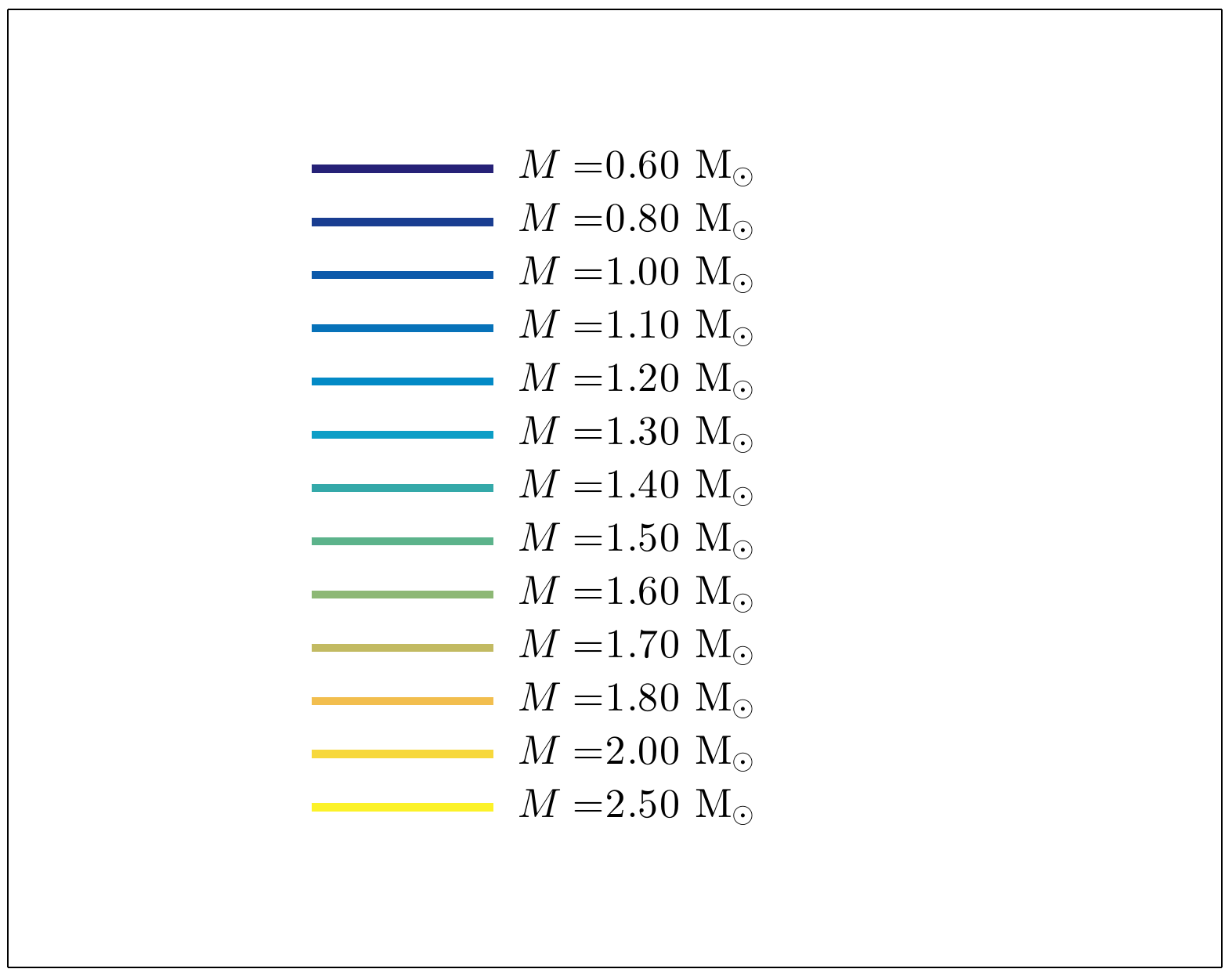}
   \includegraphics[width=8.0cm]{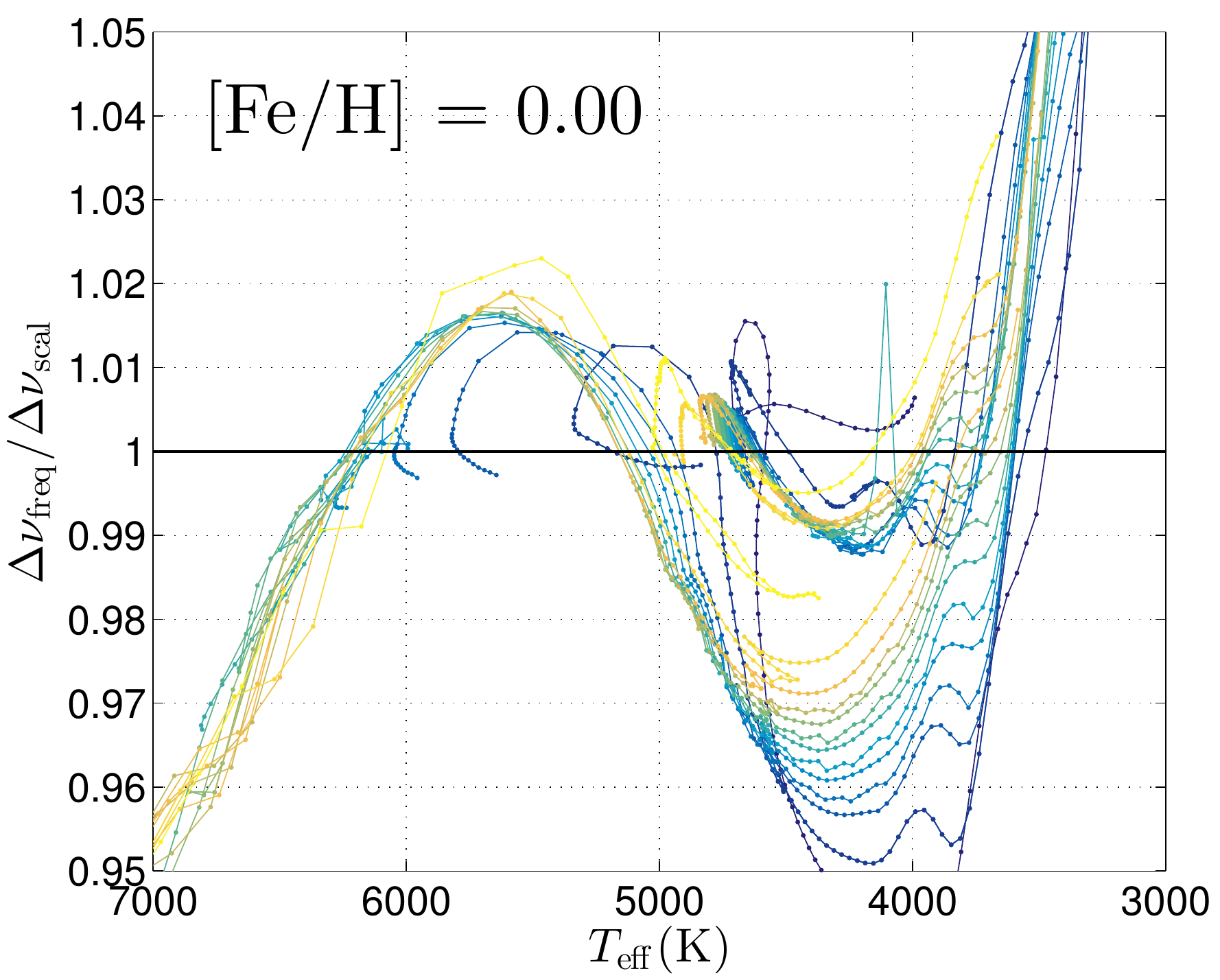}
   \caption{\textit{Upper panels}: Theoretically predicted $f_{\Delta\nu}$ for [Fe/H]=$-0.25$ and different masses as a function of $\nu_{\rm max}$ and $T_{\rm eff}$. \textit{Lower panel}: Theoretically predicted $f_{\Delta\nu}$ for [Fe/H]=$0.0$ and different masses as a function of $T_{\rm eff}$.}
             \label{fig:fdv}%
    \end{figure}

Having obtained accurate high precision dynamical measurements of mass, radius, and $T_{\rm eff}$ puts us in a position to test the accuracy of the asteroseismic predictions. For this exercise we adopt the asteroseismic measurements of $\Delta\nu$ and $\nu_{\rm max}$ by \citet{Gaulme2016}. Numbers that demonstrate the following conclusions are given in Table~\ref{table:RGdata}.
First, using the asteroseismic scaling relations in their raw form (i.e. Eqn.~\ref{eq:03} and Eqn.~\ref{eq:04} with $f_{\Delta\nu}=1$ and $f_{\nu_{\rm max}}=1$), we find that the masses and radii are significantly overestimated (for mass by 11.7\% for KIC\,7037405, 13.7\% for KIC\,9540226, and 18.9\% for KIC\,9970396, a result that has been demonstrated in many other cases by now (e.g. \citealt{Brogaard2012,Brogaard2015,Brogaard2016,Frandsen2013,Sandquist2013,Gaulme2016,Miglio2016}).
Next, using the theoretically predicted corrections to $\Delta\nu$, $f_{\Delta\nu}$ from \citet{Rodrigues2017}, reproduced in Fig.~\ref{fig:fdv} for [Fe/H]$=-0.25$, at the measured $\nu_{\rm max}$ and the masses measured in the binary analysis, and assuming $f_{\nu_{\rm max}}=1$, lower asteroseismic numbers are obtained for mass and radius, now in agreement with the dynamical estimates within the uncertainties. If the same procedure is followed while finding $f_{\Delta\nu}$ from the measured $T_{\rm eff}$ (in Fig.~\ref{fig:fdv}), a slightly larger $f_{\Delta\nu}$ (by $\sim0.001-0.002$) and thus a slightly larger mass and radius is found. As we show later in Sect.~\ref{sec:gaulme}, the masses and radii predicted by using instead the corrections to the scaling relations by \citet{Sharma2016} are quite similar to those predicted using the corrections by \citet{Rodrigues2017}.

Overall, we find that the asteroseismic scaling relations are in agreement with the dynamically measured masses and radii when they include theoretically calculated correction factors $f_{\Delta\nu}$ according to \citet{Rodrigues2017}. Thus, at least at [Fe/H]$\sim-0.25$, such a procedure seems to provide asteroseismic mass and radius estimates that are accurate to within the asteroseismic precision level of the three stars studies here ($\sim4\%$ for mass and $\sim1.5\%$ for radius).

If we trust the theoretical predictions for $f_{\Delta\nu}$, we can use the comparison between dynamical and asteroseismic masses and radii to put limits on $f_{\nu_{\rm max}}$ which we so far assumed to be 1. But before we begin this exercise, we note that the differences between dynamical and asteroseismic measures are already within the expectations according to the uncertainties when $f_{\nu_{\rm max}}=1$ and therefore we need more precise measurements for a larger sample of stars, preferably with a range of parameters including [Fe/H], in order to obtain robust indications of any potential variation of $f_{\nu_{\rm max}}$ with stellar parameters.

The numbers for $f_{\nu_{\rm max}}$ required for exact agreement between asteroseismic and dynamical measures of mass and radius are given in Table~\ref{table:RGdata}. As seen, there is no indication of a trend, since one star prefers $f_{\nu_{\rm max}}$ greater than one while the other two prefer $f_{\nu_{\rm max}}$ smaller than one. 
The predictions of the recent work by \citet{Viani2017} suggest a $T_{\rm eff}$ dependent $f_{\nu_{\rm max}}$ at a given [Fe/H], for [Fe/H]$=-0.25$ being $\sim0.990$ at 4500 K and $\sim0.999$ at $\sim 4700$ K (inferred by inverting the mass in their fig. 11). If multiplied by a factor $>1.001$ to make the latter number larger than one, which can easily be accommodated by e.g. the uncertainty in $\nu_{\rm max\odot}$ or $T_{\rm eff}$, the errors due to $f_{\nu_{\rm max}}$ suggested by \citet{Viani2017} would actually improve the self-consistency in our study. However, not only are we already working within uncertainty level, but this part of the predicted $f_{\nu_{\rm max}}$ is also not accounted for in the proposed improvement of the $\nu_{\rm max}$ scaling by \citet{Viani2017} (compare left and right panel of their fig. 11 for [Fe/H]$=-0.25$). 

We also employed the PARAM code \citet{daSilva2006,Rodrigues2014} for grid based asteroseismic modeling. The results are seen in Table~\ref{table:RGdata}. These are again in agreement with the binary measurements for mass and radius, and give in addition an age estimate corresponding to the adopted physics of the underlying stellar models. While the age estimates from PARAM are consistent with those from the eclipsing binary analysis, the predicted age difference between KIC\,7037405 and KIC\,9970396 from PARAM is much larger and much more uncertain than in the dynamical analysis, as reflected by the numbers.

%______________________________________________ 

%______________________________________________ 
   \begin{figure*}
   \centering
   \includegraphics[width=18.0cm]{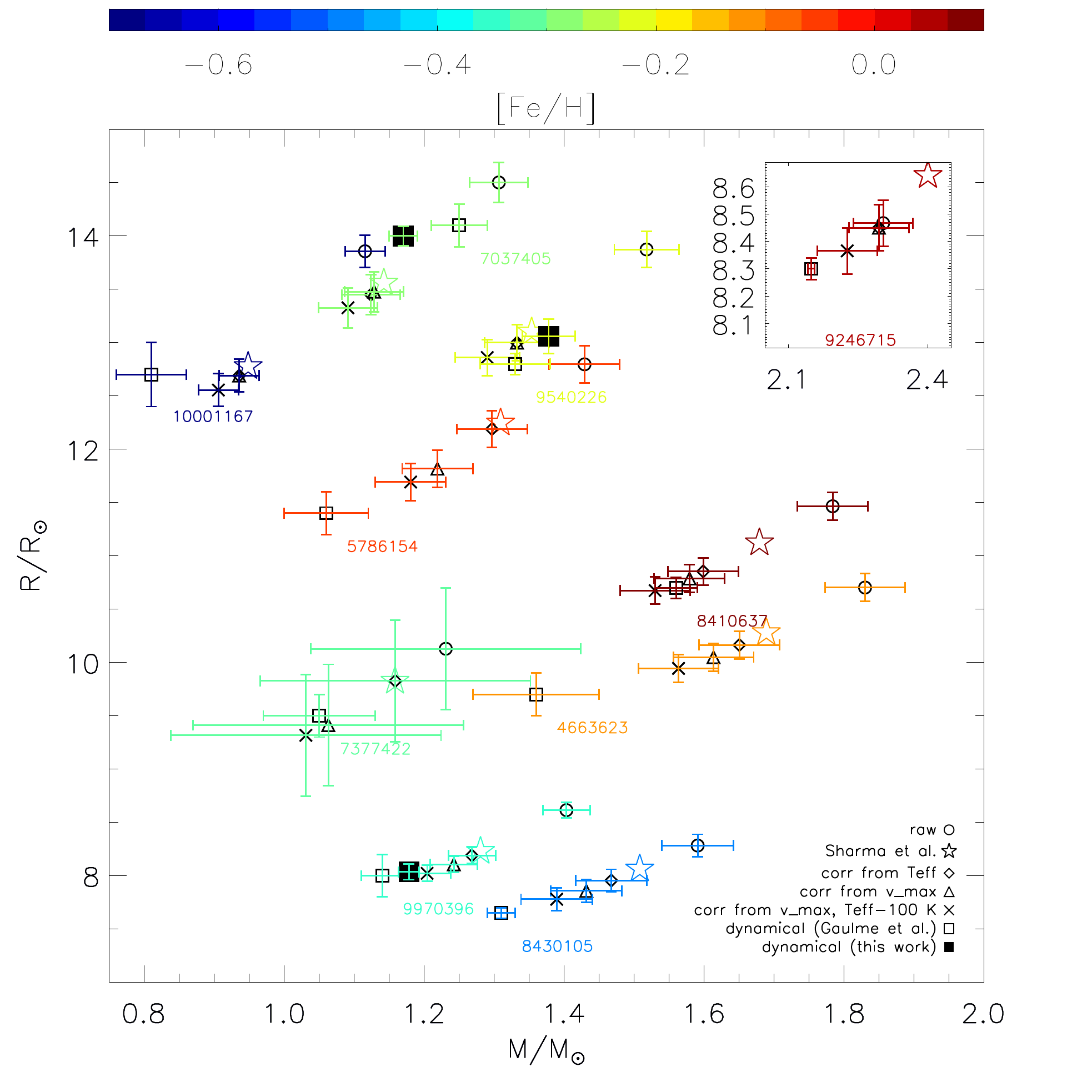}
% updated version of figure made with updates to seis_gaulme_kdv_param.pro on 181017. 
   \caption{Comparisons between dynamical and asteroseismic mass and radius estimates for 10 red giant stars in SB2 eclipsing binary systems. Open circles are estimates based on the simple asteroseismic scaling relations with $f_{\Delta \nu} = f_{\nu _{\mathrm{max}}}$ = 1. Star symbols represent scaling estimates with $f_{\Delta \nu}$ from \citet{Sharma2016}. Triangles represent scaling estimates with $f_{\Delta \nu}$ from \citet{Rodrigues2017} using $\nu_{\mathrm{max}}$ as a reference, crosses are the same, but with the observed $T_{\rm eff}$ reduced by 100 K, while diamonds are the same corrections but using $T_{\rm eff}$ as the reference. Open squares are the dynamical eclipsing binary measurements from \citet{Gaulme2016}. Big solid squares are the dynamical eclipsing binary measurements from this paper.}
             \label{fig:Gaulme}
    \end{figure*}

%______________________________________________ 
   \begin{figure*}
   \centering
   \includegraphics[width=18.0cm]{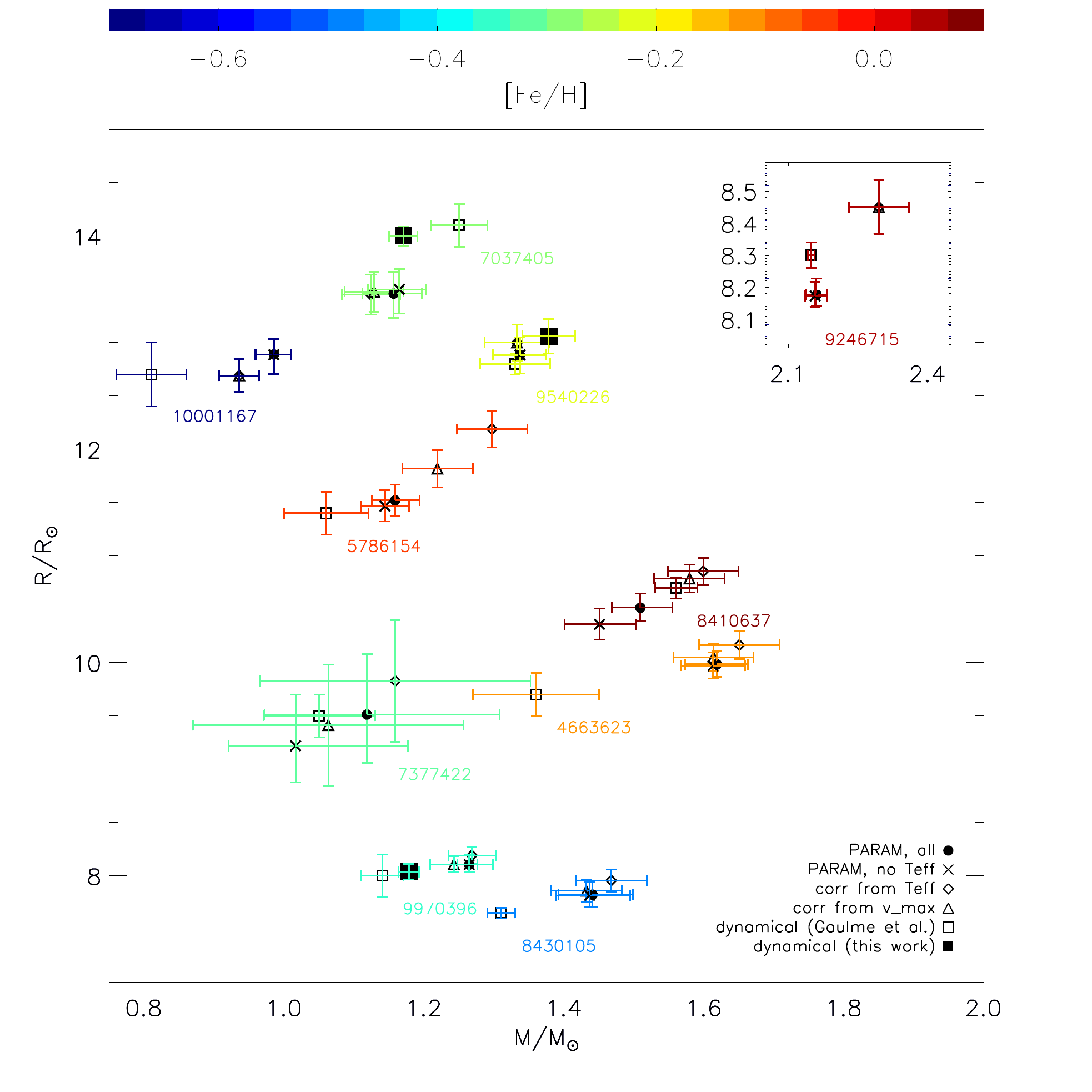}
   \caption{As Fig.~\ref{fig:Gaulme} but including PARAM grid modeling results represented by circles. Solid circles using all constraints, crosses without using $T_{\rm eff}$ as a constraint. Other symbols shown are defined as in Fig.~\ref{fig:Gaulme}. For KIC\,7037405A, KIC\,9540226A, and KIC\,9970396A we used our [Fe/H] and $T_{\rm eff}$ values.}
             \label{fig:Gaulme2}
    \end{figure*}

\section{Extension to a larger sample.}
\label{sec:gaulme}

\subsection{Comparison with other measurements.}

\citet{Gaulme2016} presented a larger sample of 10 red giant stars with masses and radii measured from eclipsing binary analysis of SB2 systems\footnote{Measurements for two of the systems were adopted by \citet{Gaulme2016} from the studies of \citet{Frandsen2013} and \citet{Rawls2016}.}, including the three measured in the present study. Fig.~\ref{fig:Gaulme} shows a mass--radius diagram of these measurements (open squares). Also plotted are our dynamical measurements for the three systems (large solid squares) we studied in the present work and asteroseimic estimates for all systems. The asteroseismic masses and radii were calculated using $\Delta \nu, \nu_{\rm max}$, and evolutionary status (RGB or Core-He-burning; they are all RGB, except for KIC\,9246715A) from \citet{Gaulme2016} and asteroseismic scaling relations with $f_{\Delta\nu}$ determined from the theoretical predictions by \citet{Rodrigues2017} using (1) $\nu_{\rm max}$ as the reference (triangles) and, alternatively (2) using $T_{\rm eff}$ as the reference (diamonds) and (3) using observed $T_{\rm eff}$ values reduced by 100 K and $\nu_{\rm max}$ as the reference (crosses). 
Star symbols represent asteroseismic scaling results using $f_{\Delta\nu}$ determined according to \citet{Sharma2016}. This was one of the cases evaluated by \citet{Gaulme2016}, and therefore allows comparison to that work. 
We return to these different choices below. We adopted $T_{\rm eff}$ from \citet{Gaulme2016}, except for the three systems of the present study where we use our $T_{\rm eff}$ estimates.

%new text:
We see in Fig.~\ref{fig:Gaulme} the same trend as reported by \citet{Gaulme2016} that the asteroseismic scaling measures without corrections are significantly overestimated. This is not really a surprise given the various investigations of theoretically expected corrections, but it is clear evidence that corrections are needed to obtain proper estimates of the properties of giant stars.
Regardless of whether we adopt corrections from \citet{Sharma2016} or \citet{Gaulme2016}, we find that employing our dynamical measures for the three stars in our present study (big solid squares in Fig.~\ref{fig:Gaulme}) improves agreement between dynamical and asteroseismic masses and radii. The only star that showed the opposite trend (dynamical estimates larger than corrected asteroseismic estimates) in the study by \citet{Gaulme2016}, KIC\,7037405A, now shows agreement between dynamical and corrected asteroseismic mass estimates. As for the other two stars in the present study, the difference to the \citet{Gaulme2016} measurements are a little more than $1\sigma$ for mass and less than $1\sigma$ for radius (their uncertainties), decreasing the tension with the asteroseismic measures for KIC\,9970396A, while for KIC\,9540226A agreement with the corrected asteroseismic estimates remains well within $1\sigma$.
%new text end in next line!
Regardless of these considerations, the general picture that remains is that for some giants the corrected asteroseismic scaling relations predict higher values than the dynamical estimates, while for other giants the dynamical and corrected asteroseismic estimates are in agreement. The corrected asteroseismic scaling estimates are in no cases significantly lower than the dynamical estimates. The comparisons of our dynamical estimates to those of \citet{Gaulme2016} (showing differences of $2\sigma, 1.25\sigma$, and $0.96\sigma$, their uncertainties) suggests that the measures of \citet{Gaulme2016} could be off by enough to account for the discrepancy with the asteroseismic scaling relations in some cases, but there is no obvious reason that this should result in the seemingly systematic nature of the differences. Therefore it makes sense to investigate the issue further, although measurements of higher precision will eventually make conclusions easier.

We can gain some insights into the cause of the tension between dynamical and asteroseismic measurements. First, by comparing the asteroseismic scaling values using either $\nu_{\rm max}$ or $T_{\rm eff}$ as the reference to obtain $f_{\Delta\nu}$ (Fig.~\ref{fig:Gaulme}; triangles and diamonds, respectively), it becomes evident that these have the largest differences for the stars that also have the largest differences between the dynamical and asteroseismic estimates. What this means is that the measured set of parameters ($\Delta \nu, \nu_{\rm max}$, $T_{\rm eff}$, [Fe/H]) for these stars are not consistent with a single stellar model in the grid used to generate the corrections when enforcing the dynamical mass. Unfortunately, this does not reveal whether the problem relates to the dynamical mass, $T_{\rm eff}$, [Fe/H], $\Delta \nu$, $\nu_{\rm max}$ or the stellar models used to obtain $f_{\Delta \nu}$.

\subsection{Potential theoretical causes}

Fig.~\ref{fig:Gaulme} shows that the apparent inconsistency between dynamical and corrected asteroseismic scaling estimates is smaller when using $\nu_{\rm max}$ to find $f_{\Delta \nu}$ rather than $T_{\rm eff}$. This suggests that a significant part of the discrepancy arises due to a mismatch between the measured $T_{\rm eff}$ and that of the model used to calculate the correction. This will cause biases in the asteroseismic scaling results by the direct $T_{\rm eff}^{3/2}$ and $T_{\rm eff}^{1/2}$ effects on mass and radius but also propagates into errors in $f_{\Delta \nu}$. The latter effect can be larger than the first but is not accounted for in uncertainty estimates using asteroseismic scaling relations. 

It is quite likely that the model $T_{\rm eff}$ scale of \citet{Rodrigues2017} is too cool, which would cause the predicted $f_{\Delta \nu}$ to be closer to one at a given measured $T_{\rm eff}$ for most of the RGB. This would lead to an overestimate of the asteroseismic masses and radii. In fact, given the significant difference between different model sets, we know that for some of the models the $T_{\rm eff}$ scale must be wrong. There are however several ways to change the model $T_{\rm eff}$ of giant stars and therefore is not possible to solve the issue without further observations. Possible, but not exhaustive possibilities include the inclusion of diffusion in the models, adjustments of the abundance pattern of the model -- in particular to take into account the [$\alpha$/Fe] variations with [Fe/H] that we know observationally is there, but are not currently taken into account in the models, potential variations in the efficiency of convection with stellar parameters implemented via the so-called mixing length parameter, and inaccuracies in the modeling of the surface boundary condition. A combination of such effects are also the most likely cause for the difference between the corrections to the scaling relations predicted by \citet{Rodrigues2017} and others who obtain the corrections by the same or very similar procedures, but using different stellar models. Indeed, in Fig.~\ref{fig:Gaulme}, the asteroseismic parameters estimated using corrections from \citet{Sharma2016} are close to, but on average slightly larger than those using the corrections by \citet{Rodrigues2017}.

KIC\,9246715A, measured by \citet{Rawls2016}, allows an observation that clearly shows the mismatch between measured and model $T_{\rm eff}$ values; This star is in the core-He-burning phase according to the asteroseismic period spacing of mixed modes \citep{Rawls2016}. However, when looking to find the predicted $f_{\Delta \nu}$ at the measured [Fe/H] = 0.0 in the lower panel of Fig.~\ref{fig:fdv}, there are no models of core-He-burning stars in the model grid used by \citet{Rodrigues2017} at the measured $T_{\rm eff}=5030\pm45$ \citep{Rawls2016} of this star. This shows that either the measured $T_{\rm eff}$ is too hot or the temperature scale of the stellar evolution models used to calculate $f_{\Delta \nu}$ is too cool. The measured $T_{\rm eff}$ is in very good agreement with core-He-burning stars in the open cluster NGC6811 that have very similar metallicity, $T_{\rm eff}$ \citep{Molenda2014} and asteroseismic parameters \citep{Arentoft2017}. In the study by \citet{Arentoft2017}, the models compared to the observations are actually on the hot side of the measured observations, illustrating that it is not unusual to have $T_{\rm eff}$ differences at the 200 K level between predictions from different models of giant stars. In Fig.~\ref{fig:Gaulme}, it can be seen that lowering the observed $T_{\rm eff}$ by 100 K, equivalent to increasing the model $T_{\rm eff}$ by 100 K provides a self-consistent solution for KIC\,9246715A. However, while a general increase to the model $T_{\rm eff}$ scale would affect the correction, $f_{\Delta \nu}$, to improve agreement in the cases where the corrected asteroseismic estimates currently seem too large, it would also introduce a systematic difference between dynamical and corrected asteroseismic measures for the three stars that we measured in the present work, in such a way that the asteroseismic masses and radii become smaller than the dynamical values. This thus seems like an unlikely solution, given that it introduces a bias for what should be the best measured stars.

Since our sample of three stars, which have the highest measurement precision of the dynamical estimates, are in agreement for $f_{\nu_{\rm max}}=1$, it is also not a likely option to shift the overall agreement for the larger sample by increasing the zero-point of $f_{\nu_{\rm max}}$ or alternatively the solar reference $\nu_{\rm max}$. Including the $f_{\nu_{\rm max}}$ correction suggested by \citet{Viani2017} would also not improve agreement, since that has a strong metallicity dependence that is not supported by the measurements.
Therefore, we need more -- and more precise -- empirical data to establish $f_{\nu_{\rm max}}$ and $f_{\Delta \nu}$.

The next natural step after using the scaling relations is to do grid based modeling. We used PARAM with the stellar parameters from \citet{Gaulme2016}, except for $T_{\rm eff}$ and [Fe/H] of our subsample, where we used our measurements. Fig.~\ref{fig:Gaulme2} shows the comparison of PARAM output masses and radii to the dynamical and scaling relation estimates. 

PARAM was run in different ways to investigate the consequences. Solid circles represent runs including the full set of observables $\Delta \nu, \nu_{\rm max}$, $T_{\rm eff}$, [Fe/H] and their uncertainties as constraints whereas crosses represent runs that did not use $T_{\rm eff}$ as a constraint. 
%The open triangles represent a case where the observed $T_{\rm eff}$ was reduced by 100 K and used as a constraint. The solid triangles represents the result of increasing the uncertainties on [Fe/H] to 0.1 dex.

As seen, for all RGB stars the grid modeling values agrees within $1\sigma$ mutual errorbars with the values from the scaling relations with $\nu_{\rm max}$ as the reference to obtain $f_{\Delta \nu}$. By comparing the different results it is evident that the $T_{\rm eff}$ sensitivity is quite different for the different stars. For example, for KIC8410637 it is clear that the grid prefers a lower $T_{\rm eff}$, perhaps even lower than it should be, given that the results without the $T_{\rm eff}$ constraint are lower than the dynamical estimates.

Overall, there is no clear trend that the PARAM results are more accurate than the corrected asteroseismic scaling results. More precise measurements of more systems are needed to investigate this.

\subsection{Potential observational causes}

Since we were not able to find an obvious theoretical reason for the apparent discrepancy between the dynamical and corrected asteroseismic masses and radii of the \citet{Gaulme2016} sample, we consider here some observational issues that could be part of the problem by causing increased or even systematic uncertainties on the measurements.

Having more stars at similar metallicities allows an inter-comparisons between systems, which suggests that $T_{\rm eff}$ could be the problematic parameter in some cases; KIC\,7377422A has a metallicity very close to that of the three stars in our sample. Although the asteroseismic estimates are very uncertain for this star, it shows the same trend as the overall sample that the corrected asteroseismic values are larger than the dynamical estimates. However, the $T_{\rm eff}$ of KIC\,7377422A is $4938\pm110$ according to \citet{Gaulme2016}, while comparisons to KIC\,7037405A and KIC\,9970396A in Fig.~\ref{fig:Gaulme} suggest that it should be between their effective temperatures of $4860-4500$ K and closer to the former than the latter. Fig.~\ref{fig:Gaulme} shows that if one adopts a $T_{\rm eff}$ lower by 100 K then nearly exact agreement between dynamical and asteroseismic numbers are reached for this star when adopting the correction from \citet{Rodrigues2017}. For KIC8410637 comparisons to very similar stars in the open cluster NGC6819 \citep{Handberg2017} suggests that the observed $T_{\rm eff}$, measured by \citet{Frandsen2013}, could be too high.

There are reasons to suspect that $T_{\rm eff}$ of the binary sample measured by \citet{Gaulme2016} could be overestimated for some stars, which would cause an overestimate of mass and radius from the asteroseismic scaling relations even if appropriate corrections are applied. The way $T_{\rm eff}$ was measured in that study ignored the continuum contribution from the secondary stars, which makes the spectral lines appear weaker, potentially mimicking a higher temperature. The stars that we re-measured taking into account the light ratio were found cooler, though only by 56, 16, and 12 K. However, the light ratio and therefore the potential overestimate of $T_{\rm eff}$ is larger for the two stars with largest discrepancy between dynamical and asteroseismic measures, KIC5786154A and KIC4663623A, than any of the three giants in our sample. We also note that the spectroscopic log$g$ values reported by \citet{Gaulme2016} for these two stars are larger than the dynamical and asteroseismic log$g$ measurements by 0.25 and 0.5 dex, suggesting potential issues with the spectroscopic analysis.

KIC10001167A can be compared to stars in the globular cluster 47 Tucanae (47\,Tuc; NGC104) given that it has similar metallicity ([Fe/H]$\sim-0.7$, compare \citealt{Gaulme2016} and \citealt{Brogaard2017}). Since the globular clusters are almost always found to be as old or older than field stars at similar metallicity, KIC10001167A would be expected to have an age equal to or younger than 47\,Tuc. A comparison to the turn-off mass of 47\,Tuc determined from the eclipsing member V69 \citep{Thompson2010}, extrapolated to the giant phase via isochrones \citep{Brogaard2017} would then suggest a mass of $\gtrsim 0.90 \rm M_{\odot}$ for KIC10001167A, about $2\sigma$ larger than the $0.81\pm0.05 \rm M_{\odot}$ measured by \citet{Gaulme2016}. Indeed, if their low mass of $0.81 \rm M_{\odot}$ is correct, the corresponding age of this star would be larger than the presently established age of the Universe. This problem could be avoided if the star is an AGB star that experienced mass-loss on the RGB. However, we also note that the $T_{\rm eff}$ measured by \citet{Gaulme2016} for KIC10001167B, the main sequence star in this binary, is very much ($\sim 6\sigma$) larger than would be expected for their measure of a $0.79\pm0.03 \rm M_{\odot}$ star on the main sequence. This suggests that the true mass of KIC10001167B is larger. Since dynamical mass estimates of the two components of a binary system correlate strongly, this also indicates that the mass of KIC10001167A is larger than the dynamical measure. Furthermore, if KIC10001167A is an AGB star then $f_{\Delta \nu}$ is $\sim1.005$ instead of $\sim0.957$ and thus the seismic mass and radius would be $\sim22$ \% larger than shown in Fig.~\ref{fig:Gaulme} thus causing a much larger disagreement with the dynamical estimates. We take this as strong indications that KIC10001167A is an RGB star with a true mass $\gtrsim 0.90 \rm M_{\odot}$ and thus in agreement with the asteroseismic estimate at the $1\sigma$ level (see Fig.~\ref{fig:Gaulme}).These considerations for the masses of KIC1001167, as well as a comparison of our dynamical measurements to those of \citet{Gaulme2016} for the three stars in common, suggest that in some cases the cause for disagreement with corrected asteroseismic scaling relations could be too low precision on the dynamical mass measurements.

\section{Summary, conclusions and outlook}
\label{sec:conclusion}

We measured precise properties of stars in the three eclipsing binary systems KIC\,7037405, KIC\,9540226, and KIC\,9970396, finding for the giant components their masses to a precision of $1.7\%, 2.8\%$, and $1.3\%$, and their radii to a precision of $0.7\%, 1.2\%$, and $0.9\%$. Using log$g$ from these measurements with the disentangled spectra of the giant components we also determined their $T_{\rm eff}$ and [Fe/H]. Combining all these precision measurements we estimated the ages of the three binary systems to be $5.3\pm0.5$, $3.1\pm0.6$, and $4.8\pm0.5$ Gyr for the adopted stellar model physics.

The dynamical measurements of the giant stars were compared to measurements of mass, radius, and age using asteroseismic scaling relations and asteroseismic grid modeling. We found that asteroseismic scaling relations without corrections to $\Delta \nu$ systematically overestimate the masses of the three red giant stars KIC\,7037405A, KIC\,9540226A, and KIC\,9970396A by 11.7\%, 13.7\%, and 18.9\%, respectively. However, by applying theoretical correction factors $f_{\Delta \nu}$ according to \citet{Rodrigues2017}, we reached general agreement between dynamical and asteroseismic mass estimates, and no indications of systematic differences at the precision level of the asteroseismic measurements.

An extension of comparisons to the larger sample of SB2 eclipsing binary stars investigated by \citet{Gaulme2016} showed a much more complicated situation, where some stars show agreement between the dynamical and corrected asteroseismic measures while others suggest significant overestimates of the asteroseismic measures. We found no simple explanation for this, but indications of several potential problems, some theoretical, others observational. The observed $T_{\rm eff}$ scale could be too hot or the model $T_{\rm eff}$ could be too cool, both of which would affect $f_{\Delta \nu}$ to incorrectly increase asteroseismic masses and radii. The neglect of the continuum contribution of the secondary components of the binary systems could also have caused an overestimate of $T_{\rm eff}$ for some stars in the study by \citet{Gaulme2016}. Comparing our dynamical measurements to those of \citet{Gaulme2016} for the three stars in common, and comparing their dynamical mass of KIC10001167A to that of stars in the globular cluster 47\,Tuc suggests that in some cases the precision on the dynamical measurements could be part of the problem.
 
We found no indication that $f_{\nu_{\rm max}}$ should be different from 1 from our sample of three stars. These have higher precision on the dynamical measurements than the larger sample, which suggests that it is also not a viable option to shift the overall agreement for the larger sample by increasing the zero-point of $f_{\nu_{\rm max}}$ or alternatively the solar $\nu_{\rm max}$.

In order to make progress and establish and improve the accuracy level of asteroseismology of giant stars across mass, radius and metallicity, we need to (1) improve the precision of the dynamical parameters of the known sample and (2) increase the sample with precision measurements significantly to span a large range of stellar parameters. Both can be achieved by detailed observations and analysis of known {\it Kepler} targets as in the present paper, extended also to new targets found by K2 and by the upcoming surveys TESS and PLATO. Once Gaia \citep{Gaia2016} delivers accurate distances to these systems the observed $T_{\rm eff}$ scale can be constrained for bright targets to a level where the stellar model temperature scale can also be challenged. With enough observations it will be possible to reach a S/N level for the separated secondary components that would allow direct $T_{\rm eff}$ estimates of these. Since stellar models respond quite differently to changes in model physics for the main sequence and red giant phases, this will provide means to distinguish between potential ways of adjusting the model $T_{\rm eff}$ scale.

In the longer term the development of asteroseismology will be to make use of individual mode frequencies instead of just the average asteroseismic parameters $\Delta \nu$ and $\nu_{\rm max}$. That should allow increased precision of the measurements \citep{Miglio2017}, but calibration stars with precise, accurate and independent measurements will still be needed to establish also accuracy. Therefore, we strongly encourage continued efforts to find and measure as many detached eclipsing binary stars with potential to do asteroseismology as possible.

%______________________________________________ 
%   \begin{figure}
%   \centering
%   \includegraphics[width=8.6cm]{mass_legend2.pdf}
%   \includegraphics[width=8.6cm]{nmax_5.pdf}   
%   \includegraphics[width=8.6cm]{T_5.pdf}
%   \caption{Theoretically predicted $f_{\Delta\nu}$ for [Fe/H]=$0.0$ and different masses as a function of $T_{\rm eff}$.}
%             \label{fig:fdvfeh0}%
%    \end{figure}

\section*{Acknowledgements}

We thank the anonymous referee for useful comments that helped improve the paper.

Based in part on observations made with the Nordic Optical Telescope, operated by the Nordic Optical Telescope Scientific Association at the Observatorio del Roque de los Muchachos, La Palma, Spain, of the Instituto de Astrofisica de Canarias.

Funding for the Stellar Astrophysics Centre is provided by The Danish National Research Foundation (Grant DNRF106). The research was supported by the ASTERISK project (ASTERoseismic Investigations with SONG and Kepler) funded by the European Research Council (Grant agreement no.: 267864).

AM, GRD, KB and WJC acknowledge the support of the UK Science and Technology Facilities Council (STFC).

This paper includes data collected by the Kepler mission. Funding for the Kepler mission is provided by the NASA Science Mission directorate.

Some of the data presented in this paper were obtained from the Mikulski Archive for Space Telescopes (MAST). STScI is operated by the Association of Universities for Research in Astronomy, Inc., under NASA contract NAS5-26555. Support for MAST for non-HST data is provided by the NASA Office of Space Science via grant NNX09AF08G and by other grants and contracts.

%%%%%%%%%%%%%%%%%%%%%%%%%%%%%%%%%%%%%%%%%%%%%%%%%%



\bibliographystyle{mnras}
\bibliography{brogaard} % if your bibtex file is called example.bib

%\begin{thebibliography}{99}
%\end{thebibliography}

%%%%%%%%%%%%%%%%%%%%%%%%%%%%%%%%%%%%%%%%%%%%%%%%%%

%%%%%%%%%%%%%%%%% APPENDICES %%%%%%%%%%%%%%%%%%%%%

\appendix

\section{Tables with RV measurements}

\begin{table}
\centering
%\small
\caption{Individual RV measurements for KIC\,7037405}
    \begin{tabular}{lrr}
\hline
\hline
BJD & $\rm RV_G$ (km/s) & $\rm RV_{MS}$ (km/s) \\
\hline
56752.66993700 & -50.42(19) & -27.75(58)\\ 
56784.53203129 & -19.88(17) & -59.34(62)\\
56801.53711338 & -11.70(18) & -67.97(19)\\
56928.44124350 & -59.33(19) & -18.19(54)\\
57118.67783621 & -57.23(23) & -18.80(46)\\
57121.70507447 & -58.20(21) & -19.34(80)\\ 
57138.70040506 & -58.78(19) & -17.52(62)\\
57143.66574144 & -58.53(15) & -17.84(29)\\
57207.52175657 & -13.84(17) & -65.46(61)\\
57209.61205011 & -12.94(15) & -66.78(49)\\
57211.69399493 & -12.24(22) & -67.53(51)\\
57226.43521905 & -13.65(17) & -66.41(33)\\
57229.62424435 & -14.91(15) & -65.30(71)\\
57508.63399317 & -51.17(19) & -27.1(1.5)\\ 
57527.55711169 & -56.75(21) & -21.79(68)\\ 
57541.59971017 & -58.90(20) & -18.12(52)\\
57546.55162312 & -59.31(17) & -18.0(1.1)\\
57564.65749689 & -57.64(18) & -19.83(68)\\ 
57564.67409692 & -57.58(15) & -20.12(25)\\ 
57573.67390413 & -54.56(15) & -23.14(24)\\ 
57584.64858487 & -48.01(21) & -30.19(46)\\ 
57605.57042742 & -27.33(21) & -50.78(53)\\
57647.53391490 & -16.58(16) & -62.0(1.6)\\
\hline
    \end{tabular} 
\label{tab:rv7037405}
\end{table}

\begin{table}
\centering
%\small
\caption{Individual RV measurements for KIC\,9540226}
    \begin{tabular}{lrr}
\hline
\hline
BJD & $\rm RV_G$ (km/s) & $\rm RV_{MS}$ (km/s) \\
\hline
55714.63603200 & -21.17(14) &  0.38(43)\\
55714.65915553 & -21.41(11) & -1.2(1.9)\\
55714.67883338 & -21.45(10) & -0.15(48)\\
55733.64458081 & -25.61(11) &  8.2(1.4)\\
55749.56827697 & -26.49(10) &  8.48(75)\\
55762.67188540 & -25.65(12) &  8.5(1.6)\\
55795.52703045 & -18.42(11) & (...)\\
55810.50443762 &  -9.68(17) & (...)\\
55834.48647535 &  16.46(12) & -50.0(1.3)\\
55844.36927615 &  18.95(14) & -53.3(1.4)\\
55765.49842664 & -25.88(12) & 8.1(1.3)\\
55783.50678223 & -22.49(09) & 3.0(1.3)\\
55872.38744637 & -12.11(27) & (...)\\
55884.34338525 & -19.27(19) & -1.8(5.4)\\
55884.35785526 & -19.29(16) & -1.7(2.7)\\
55889.33262462 & -21.23(18) & (...)\\
55889.34683006 & -21.26(20) & (...)\\
55889.36715003 & -21.24(18) & (...)\\
55990.74296737 &  -5.04(12) & (...)\\
56106.49571278 & -26.51(09) & 8.9(1.8)\\
56106.51712882 & -26.51(11) & 9.25(61)\\
56126.65890198 & -24.14(12) & 5.9(2.1)\\
56126.68031735 & -24.17(14) & 3.8(2.8)\\
56132.52595636 & -23.02(14) & 4.9(1.5)\\
56132.54782967 & -22.96(15) & 3.8(3.5)\\
56136.61201135 & -21.59(12) & -0.0(1.8)\\
56136.63342660 & -21.61(16) & 1.8(1.0)\\
56139.45074171 & -20.96(13) & 2.0(1.7)\\
56148.50573531 & -17.42(11) & (...)\\
56148.52715013 & -17.40(12) & (...)\\
56158.49770203 & -11.55(14) & (...)\\
56158.51911708 & -11.51(09) & (...)\\
56176.44639754 &   6.06(10) & -37.08(67)\\
56182.41124337 &  13.48(10) & -47.12(81)\\
56184.56775140 &  15.84(14) & -50.1(1.5)\\
56184.59032381 &  15.95(14) & -49.3(2.6)\\
56195.53024667 &  18.75(17) & -54.4(2.0)\\
56506.41176785 & -13.45(10) & (...)\\
56506.43318172 & -13.52(11) & (...)\\
56510.50372774 & -10.69(17) & (...)\\
56518.39330906 &  -4.05(14) & (...)\\
56519.46756916 &  -2.96(09) & (...)\\
\hline
    \end{tabular} 
\label{tab:rv9540226}
\end{table}

\begin{table}
\centering
%\small
\caption{Individual RV measurements for KIC\,9970396}
    \begin{tabular}{lrr}
\hline
\hline
BJD & $\rm RV_G$ (km/s) & $\rm RV_{MS}$ (km/s) \\
\hline
56411.60261064 &    7.60(15) & -43.62(21)\\
56784.55828128 &  -33.99(15) & 5.50(74)\\
56801.56404039 &  -31.83(16) &  2.74(30)\\
56928.41286933 &  -10.25(18) & -22.79(57)\\
57118.70592049 &    7.91(17) & -43.01(41)\\
57121.72999201 &    6.85(17) & -42.62(69)\\
57143.69272636 &   -0.46(15) & -32.59(29)\\
57207.54746093 &  -26.98(15) & -2.38(39)\\
57209.63741296 &  -27.66(17) & -1.76(46)\\
57211.71904877 &  -28.26(21) & -0.66(43)\\
57226.46072639 &  -31.52(16) & 2.79(37)\\
57229.64921341 &  -32.17(18) & 3.40(45)\\
57502.67601494 &  -32.78(18) & 4.22(37)\\
57526.56249922 &  -24.44(15) & -5.84(22)\\
57528.54475853 &  -23.33(18) & -7.36(38)\\
57557.58915972 &   -2.47(15) & -31.45(46)\\
57560.55009731 &   -0.41(16) & -34.00(35)\\
57564.69442973 &    2.14(16) & -37.00(62)\\
57571.66870647 &    5.46(17) & -40.80(28)\\
57573.70065709 &    6.18(18) & -41.84(15)\\
57584.62157623 &    7.72(19) & -43.92(45)\\
57605.59761277 &    2.69(19) & -37.38(44)\\
\hline
    \end{tabular} 
\label{tab:rv9970396}
\end{table}


%%%%%%%%%%%%%%%%%%%%%%%%%%%%%%%%%%%%%%%%%%%%%%%%%%

% Don't change these lines
\bsp	% typesetting comment
\label{lastpage}
\end{document}